\documentclass[
    a4paper,
    twocolumn,
    superscriptaddress]{revtex4-1}
\usepackage[utf8]{inputenc}
\usepackage[colorinlistoftodos, color=green!40, prependcaption]{todonotes}
\usepackage{amsmath}
\usepackage{xcolor}
\usepackage{times}
\usepackage{mathptmx}
\usepackage{newtxmath}
\usepackage{graphicx}
\usepackage[T1]{fontenc}
\usepackage{float}
\usepackage[pdftex, pdftitle={Potential benefits of delaying the second mRNA COVID-19 vaccine dose}, pdfauthor={B. F. Maier et al.}]{hyperref} 
\begin{document}

\title{Potential benefits of delaying the second mRNA COVID-19 vaccine dose}

\author{B.~F.~Maier}
    \email[Correspondence email address: ]{bfmaier@physik.hu-berlin.de}
    \affiliation{Institute for Theoretical Biology and Integrated Research Institute for the Life-Sciences, Humboldt University of Berlin, Germany}
    
\author{A.~Burdinski}
    \affiliation{Institute for Theoretical Biology and Integrated Research Institute for the Life-Sciences, Humboldt University of Berlin, Germany}
    
\author{A.~H.~Rose}
    \affiliation{Institute for Theoretical Biology and Integrated Research Institute for the Life-Sciences, Humboldt University of Berlin, Germany}
    
\author{F.~Schlosser}
     \affiliation{Institute for Theoretical Biology and Integrated Research Institute for the Life-Sciences, Humboldt University of Berlin, Germany}
     
\author{D.~Hinrichs}
     \affiliation{Institute for Theoretical Biology and Integrated Research Institute for the Life-Sciences, Humboldt University of Berlin, Germany}
     
\author{C.~Betsch}
    \affiliation{University of Erfurt, Germany}
\author{L.~Korn}
    \affiliation{University of Erfurt, Germany}
\author{P.~Sprengholz}
    \affiliation{University of Erfurt, Germany}
\author{M.~Meyer-Hermann}
    \affiliation{Department of Systems Immunology and Braunschweig Integrated Centre of Systems Biology (BRICS), Helmholtz Centre for Infection Research, Braunschweig, Germany}
    \affiliation{Institute for Biochemistry, Biotechnology and Bioinformatics, Technische Universität Braunschweig, Braunschweig, Germany}
\author{T.~Mitra}
    \affiliation{Department of Systems Immunology and Braunschweig Integrated Centre of Systems Biology (BRICS), Helmholtz Centre for Infection Research, Braunschweig, Germany}
\author{K.~Lauterbach}
     \affiliation{Institute of Health Economics and Clinical Epidemiology, University of Cologne}
     \affiliation{Department of Health Policy and Management, Harvard School of Public Health, Boston}
\author{D.~Brockmann}
     \affiliation{Institute for Theoretical Biology and Integrated Research Institute for the Life-Sciences, Humboldt University of Berlin, Germany}

\date{\today} 

\begin{abstract}
Vaccination against COVID-19 with the recently approved mRNA vaccines BNT162b2 (BioNTech/Pfizer) and mRNA-1273 (Moderna) is currently underway in a large number of countries. However, high incidence rates and rapidly spreading SARS-CoV-2 variants are concerning. In combination with acute supply deficits in Europe in early 2021, the question arises of whether stretching the vaccine, for instance by delaying the second dose, can make a significant contribution to preventing deaths, despite associated risks such as lower vaccine efficacy, the potential emergence of escape mutants, enhancement, waning immunity, reduced social acceptance of off-label vaccination, and liability shifts. A quantitative epidemiological assessment of risks and benefits of non-standard vaccination protocols remains elusive. To clarify the situation and to provide a quantitative epidemiological foundation we develop a stochastic epidemiological model that integrates specific vaccine rollout protocols into a risk-group structured infectious disease dynamical model. Using the situation and conditions in Germany as a reference system, we show that delaying the second vaccine dose is expected to prevent COVID-19 deaths in the four to five digit range, should the incidence resurge in the first six months of 2021. We show that this considerable public health benefit relies on the fact that both mRNA vaccines provide substantial protection against severe COVID-19 and death beginning 12 to 14 days after the first dose. The model predicts that the benefits of protocol change are attenuated should vaccine compliance decrease substantially. To quantify the impact of protocol change on vaccination adherence we performed a large-scale online survey. We find that, in Germany, changing vaccination protocols may lead only to small reductions in vaccination intention depending on liability issues associated with postponing the second dose. In sum, we therefore expect the benefits of a strategy change to remain substantial and stable.
\end{abstract}


\maketitle

\section{Introduction}
\label{sec:intro}
Large-scale vaccine rollouts against coronavirus disease 2019 (COVID-19)
began on a world-wide scale at the end of 2020. Because of variable
vaccine supplies, distribution and delivery protocols, national policies
and vaccination strategies, the rate at which vaccines are administered
varies substantially in different countries. For instance, Israel had
reached a national vaccine coverage of more than 28\% in the general
population by Feb 16, 2021 and started vaccinating adolescents on Jan
23, 2021 \cite{israel_ministry_of_health_corona_2021}. By the same date, many European countries had only vaccinated
a fraction of the population in the single digit percent range
\cite{commissario_straordinario_per_lemergenza_covid-19_-_presidenza_del_consiglio_dei_ministri_covid-19_2021,bundesministerium__fur__soziales__gesundheit__pflege__und__konsumentenschutz_corona-schutzimpfung_2021,sante_publique_france_donnees_2021,robert_koch_institute_digitales_2021}.
In Germany, for example, by Feb 15, 2021 only 3\% of the population had
received the first dose and at the time of writing the vaccination
progresses at approximately 120,000-150,000 persons per day
\cite{robert_koch_institute_digitales_2021}. Currently approved
mRNA vaccines BNT162b2 (BioNTech/Pfizer) and mRNA-1273 (Moderna) are
reported to be approximately 95\% efficacious against COVID-19 disease
after receiving two doses \cite{polack_safety_2020,voysey_safety_2021}.
Efficacy against severe COVID-19 and death is projected to be close to
100\%, whereas overall risk reduction of infection by the causing agent
severe acute respiratory syndrome coronavirus type 2 (SARS-CoV-2) is
found to be less considerable
\cite{polack_safety_2020,voysey_safety_2021,chodcik_effectiveness_2021}.
Therefore, and because vaccination rates are limited by supply and
infrastructural challenges, vaccination strategies typically prioritize
the population at risk, i.e.~the elderly, people with comorbidities, and
people with high risk of exposure
\cite{vygen-bonnet_beschluss_2020,bullard_predicting_2020}. The public
health priority during this early phase is therefore the direct and
effective reduction of severe COVID-19 cases and deaths in the high-risk
groups as opposed to reducing transmission frequency in the general
population.

\begin{figure*}
    \centering
    \includegraphics[width=\textwidth]{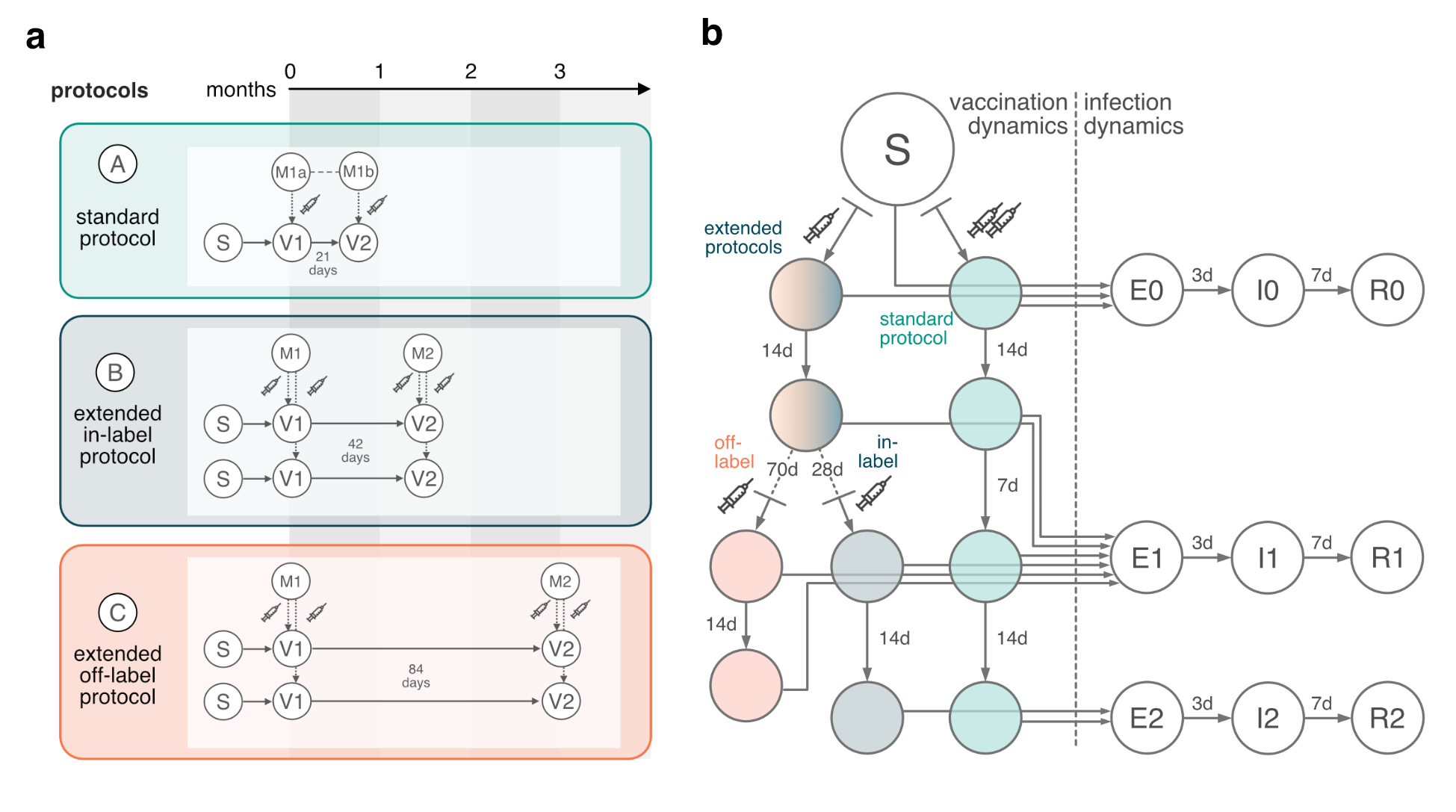}
    \caption{Vaccine administration protocols and simplified epidemiological model architecture. (a) In the standard protocol, every individual that receives a first dose is registered to receive a second dose 21 days later. The first dose is administered immediately, the second dose is reserved and appropriately stored for 21 days. In the extended in-label protocol, the second dose is not stored, but administered to a second individual instead. Both individuals will receive a second dose 42 days later (to stay within the time specifications of the manufacturers). In the extended off-label protocol, both individuals receive the second dose about 12 weeks (84 days) later, i.e.~in the consecutive quarter (or earlier, if supply is sufficient). (b) We employ a modified susceptible-exposed-infectious-removed model with a latency period of 3.2 days and an infectious period of 6.7 days. The population consists of two risk groups with varying infection fatality rate---shown here is one layer for a single risk group. Risk group layers are coupled by infection terms that are proportional to the average number of contacts between individuals of the respective groups. Because the mRNA vaccines do not primarily protect against infection with SARS-CoV-2 but rather against COVID-19, all vaccinated individuals are classified as susceptible, associated with decreased susceptibility for protection levels 1 and 2, respectively. For increasing vaccine protection level, we assume increased efficacy against COVID-19 and COVID-19-related death. After receiving the first dose, individuals remain unprotected for approximately 14 days. We conservatively assume that initial protection remains constant and is not boosted for individuals that receive the second dose after 12 weeks (84 days). Note that we compute the number of deaths and symptomatic cases as the respective fractions of the ``removed'' compartment.}
    \label{fig:1}
\end{figure*}

According to the World Health Organization's Strategic Advisory Group of
Experts on Immunization (SAGE) as well as manufacturers'
recommendations, individuals that receive the first dose of the
respective vaccine should receive the second dose after approximately 21
or 28 days, but no later than 42 days after the first dose
\cite{strategic_advisory_group_of_experts_on_immunization_sage_interim_2021,strategic_advisory_group_of_experts_on_immunization_sage_interim_2021-1}.
To comply with these recommendations, the current standard vaccine
administration protocol that is applied in the majority of European
countries provides the first dose immediately, and the second dose is
withheld and stored appropriately to be used on the same individual 21
to 28 days later (see Fig.~\ref{fig:1}a). Reserving the second dose,
however, implies that the stock of available doses is depleted twice as
fast as the number of people with first dose protection increases.
Studies indicate that a substantial amount of protection is already
delivered by the first dose
\cite{polack_safety_2020,voysey_safety_2021,chodcik_effectiveness_2021}.
After 12-14 days of receiving the initial dose, efficacy was
approximately 89\% and 92\% for BNT162b2 and mRNA-1273, respectively
\cite{polack_safety_2020,voysey_safety_2021,
strategic_advisory_group_of_experts_on_immunization_sage_interim_2021,strategic_advisory_group_of_experts_on_immunization_sage_interim_2021-1}.
Supported by additional
epidemiological data from Israel, this implies that the first dose
provides almost complete protection against severe COVID-19 and death
\cite{polack_safety_2020,voysey_safety_2021,chodcik_effectiveness_2021,jabal_impact_2021,cohen_one-dose_2021,amit_early_2021,public_health_england_annex_2021}.
Consequently, a situation of severe vaccine scarcity could potentially
be improved by instead of reserving the second dose on an individual
basis, administering it to a second person on the same date, thereby
doubling the initial vaccination rate and substantially reducing the
time required to protect the high-risk population (see
Fig.~\ref{fig:1}a). Naturally, this would delay the administration of
the second dose if vaccine supply is limited. Such a protocol is being
followed in the UK, for instance, where the Joint Committee on
Vaccination explicitly recommends to prioritize the first-dose
vaccination of as many high-risk individuals as possible over a timely
administration of the second dose
\cite{department_of_health_and_social_care_priority_2020}.

In light of the current situation including concerns about the spread of new virus variants with substantially higher transmission probabilities, discussions of whether the benefits of these alternative vaccination protocols outweigh their risks are highly disputed and controversial \cite{department_of_health_and_social_care_priority_2020,noauthor_science_2021,iacobucci_covid-19_2021,sewell_revisiting_2021}.

Many European countries face a shortage of available mRNA vaccine doses within the first quarter of 2021, a situation that is likely to improve in the second quarter when the number of delivered doses is expected to be significantly greater. Therefore, a systematic delay of the second dose administration could potentially save many lives by maximizing vaccination speed and scope. However, so far only few systematic analyses have been performed that compare potential epidemiological outcomes of alternative strategies to the standard protocol \cite{wang_impacts_2021}. In what way and under which conditions alternative protocols affect the expected number of severe COVID-19 cases and deaths remains yet unknown but could significantly guide policy makers to make informed decisions.

Given this and the substantial protection against severe COVID-19 and
death provided by a single dose, it might be beneficial to delay the
second dose by 12 weeks, reaching twice as many people in the first
quarter of 2021 compared to the standard protocol. However, exceeding
the recommended interval between the first and second doses comes with
the hypothetical risk of decreased vaccine efficacy, as well as the
potential loss of manufacturer liability when administering vaccines by
an off-label procedure. A further risk is a potential reduction in the
population’s willingness to get vaccinated when the vaccines are used on
off-label protocols.

Alternatively, and in order to comply with the manufacturers' recommendations, the second dose can be delayed up to 42 days in an extended in-label protocol
\cite{european_medicines_agency_clarification_2021,strategic_advisory_group_of_experts_on_immunization_sage_interim_2021,strategic_advisory_group_of_experts_on_immunization_sage_interim_2021-1}.
As contracts over vaccine dose deliveries are usually signed per quarter (three months), the same number of individuals can be vaccinated per quarter for both the standard and extended in-label protocol. In the extended in-label protocol, however, all high-risk individuals will have received the second dose after about six weeks which is within the originally approved time frame.

Here we employ a stochastic computational model to capture the epidemiological impact of changing the standard protocol and implementing in- and off-label extended protocols instead. A particular emphasis is placed on quantifying the public health benefits or drawbacks of these alternative strategies in terms of the expected number of saved lives and avoided COVID-19 cases in the high-risk population. The model is designed to shed light onto the factors that may yield benefits of alternative protocols and identify conditions for which benefits become drawbacks. In particular, we investigate how a potentially negative impact of an alternative vaccination protocol, especially off-label use, on vaccination adherence may neutralize the benefits expected by faster coverage. The model compares three categorically different pandemic scenarios for the first 6 months in 2021.

We find that due to the high efficacy of the vaccines with respect to death from COVID-19, a substantial reduction in fatal outcomes is expected for all extended protocols and all pandemic scenarios. However, should a change in protocol impact vaccine adherence substantially, our model indicates that these benefits are neutralized and become disadvantageous.
 
\section{Model and methods}
\label{sec:model}
In order to assess the impact and potential benefits of different vaccination protocols, the model requires detailed data on vaccine supplies which can vary between countries. We therefore investigate two different supply timelines and scenarios. Both scenarios are derived from the particular situation in Germany (population of the order of 80 million).

In both scenarios we assume that in the first quarter of 2021, 14 million doses of mRNA vaccine are available and are delivered at a constant rate over the whole quarter. A parsimonious ``extreme scarcity'' scenario focuses on the general effects expected for different vaccination protocols and assumes the same number of doses for each quarter in 2021, so after 6 months a total of 14 million individuals can be vaccinated with two doses. The results obtained for these values can be used as a guide for the expected effects in other countries and as a reference scenario.

The second scenario is based on actual estimations on mRNA vaccine availability in Germany \cite{federal_ministry_of_health_of_germany_informationen_2021}. Here, 46 million doses are expected to be available in the second quarter of 2021. This would amount to a total of 30 million people that could receive two doses of the discussed vaccines in the first half of 2021. We explicitly ignore the influence of Adenovirus vaccines that might become available additionally, in order to solely focus on the impact made by delaying the second mRNA vaccine dose.

Analyzing both scenarios enables us to compare the benefits of protocol changes for both extreme scarcity as well as moderate scarcity of vaccine supply.

We developed and implemented a risk-group-structured stochastic susceptible-exposed-infectious-removed (SEIR) model to capture the combined dynamics of the time-course of the pandemic and the impact of vaccination (see Fig.~\ref{fig:1}b) \cite{anderson_infectious_2010,keeling_modeling_2011} with a uniform average latent period of 3.2 days and an infectious period of 6.7 days \cite{khailaie_development_2021}. Using a coarse-grain approximation, we divide the population into two groups: high-risk (which we define as having increased risk of death from infection) and low-risk (low risk of death from infection). We assume that the high-risk group comprises approx. 20\% of the population, requiring swift and high-priority vaccination (see \cite{vygen-bonnet_beschluss_2020} and Supplementary Information (SI)). We model the average number of epidemiologically relevant contacts between individuals of one risk group to another risk group based on results by the POLYMOD study, using the 65+ age group as a proxy for the ``high-risk'' group and the <65 age group for ``low-risk'' \cite{mossong_social_2008,funk_socialmixr_2020}. While the official definition for ``high-risk'' begins at age 70, individuals with co-morbidities are classified as ``high-risk'' as well, many of which belong to the age group 65-70 \cite{mueller_why_2020}. Based on empirical results, we expect that approx. 75\%-80\% (i.e.~12 million people) of the high-risk group are willing to get vaccinated (see Fig.~\ref{fig:4}a). Additionally, we assume that 7 million individuals of the low-risk group are workers in the health care sector and consequently are vaccinated with high priority in concurrence with the high-risk population \cite{vygen-bonnet_beschluss_2020}.

We calibrate the model based on the reported daily incidence and
COVID-19-related deaths in Germany in 2020, yielding plausible values of
infection fatality rates (IFRs) of 6.25\% for the high-risk and 0.028\%
for the low-risk group \cite{levin_assessing_2020}, further details are
provided in the SI. We find an in-group base reproduction number of $R_0
= 2.2$ for the low-risk group which we use to scale cross-group
transmission rates accordingly (see SI). Additionally, we set an 83\% probability of infected individuals to display symptoms of COVID-19 \cite{byambasuren_estimating_2020}.

Vaccination protection is primarily aimed at preventing disease and
reducing COVID-19 deaths. Vaccines are expected to be less effective
with respect to preventing SARS-CoV-2 infection
\cite{polack_safety_2020,voysey_safety_2021,chodcik_effectiveness_2021,amit_early_2021,public_health_england_annex_2021}.
A vaccinated person is almost fully protected against COVID-19 related
death approx.~14 days after receiving the first dose
\cite{chodcik_effectiveness_2021}. In the model we therefore set
efficacy against death to be 99.9\% after the first dose and increase it
to 99.99\% after two doses. We further expect that the efficacy against
COVID-19 is 90\% after the first dose as compared to 95\% after
receiving the second dose \cite{polack_safety_2020,voysey_safety_2021}. We implement values
of a 50\% reduction in susceptibility approx.~14 days after receiving
the first dose \cite{chodcik_effectiveness_2021}, and assume an increase
of susceptibility reduction to 60\% after the second dose if
administered according to the standard protocol, i.e.~21 days plus 14
days after the first dose. We assume that vaccination following the
extended off-label protocol provides overall less protection against the
disease compared to the standard and extended in-label protocols,
respectively. These assumptions are conservative and deliberately
underestimate the effectiveness of the second dose when administered
with a 12 week delay. Fixed at a lower level, we additionally assume
that vaccination protection does not decrease significantly in the first
12 weeks after receiving the first dose, and that, after 12 weeks, the
second dose only maintains this protection rather than increasing it.

The course of the pandemic over the next six months cannot be predicted due to the complexity of the process and feedback effects between pandemic dynamics, population behavioral changes, and policy measures \cite{ioannidis_forecasting_2020,prasse_fundamental_2020}. An important empirical finding of the analysis of the pandemic in Germany and other countries, however, is that while the epidemic unfolds, a dynamic equilibrium is achieved in which the natural force of infection and response measures balance, causing the effective reproduction number $R(t)$ to continuously fluctuate around the system-critical value $R(t) \approx 1$, a phenomenon known as self-organized criticality that has been observed in disease dynamical systems \cite{funk_spread_2009} and other complex dynamical systems \cite{bak_self-organized_1987,bak_punctuated_1993}. The statistical properties of $R(t)$ fluctuations can therefore be used to generate long-term forecasts from quasi-stationary and temporally homogeneous driven stochastic processes that resemble empirical incidence curves.

\begin{figure*}
    \centering
    \includegraphics[width=\textwidth]{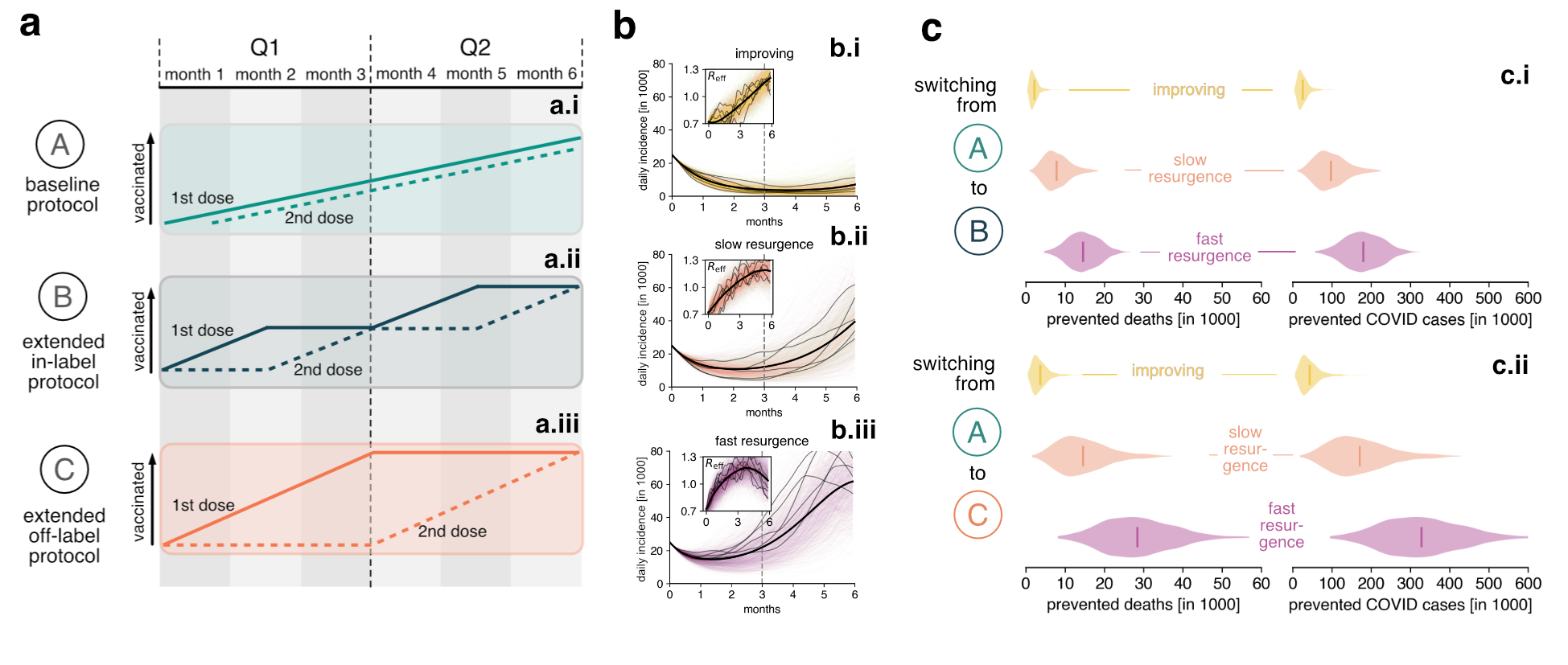}
    \caption{Model results for extreme scarcity of vaccine supply (14 million doses in the first quarter (Q1) and 14 million doses in the second quarter (Q2)). (a) Schematic of vaccine dose distribution to high priority individuals (12 million individuals in high-risk group and 7 million individuals in low-risk group). Note that for the standard protocol (a.i) and extended in-label protocol (a.ii) the same number of people can receive the first dose in Q1 (7 million people). However, for the extended in-label protocol, all individuals received their first dose within 6 weeks instead of 12 weeks. For the extended off-label protocol, the number of people that received the first dose in Q1 doubles. (b) We simulate three possible scenarios of the epidemic situation in the first months during the vaccine rollout. For each scenario, 1000 independent realizations were computed. Details of the model are provided in the SI. Shown here are respective incidences for a scenario in which no vaccines would be distributed, each lightly colored line represents a single simulation, grey lines are illustrative examples of single simulations and black lines represent the mean over all simulations. Insets show the respective effective reproduction number. (b.i) ``Improving'' scenario: incidence decreases and remains on a low level during rollout. (b.ii) ``Slow resurgence'':  incidence rises after an initial decrease to cause a low third wave in the second quarter. (b.iii) ``Fast resurgence'': incidence quickly resurges to cause large incidences in the second quarter. (c) Prevented high-risk group deaths and symptomatic cases when switching from baseline to extended protocols (refer to Tab.~\ref{tab:1})). (c.i) Following the extended in-label protocol instead of the standard protocol will save lives in the four- to low five-digit range. (c.ii) The number of saved lives and prevented symptomatic cases almost doubles when changing to the extended off-label protocol instead.}
    \label{fig:2}
\end{figure*}

With regard to the development of the pandemic in 2021, we therefore
address three different basic scenarios based on transmission rates that
follow a stochastic process (see Fig.~\ref{fig:2}b and SI). In the
``improving'' scenario, the expected incidence continues to decrease
over the first and second quarter of 2021, reaching a low, approximately
constant level. In the ``slow resurgence'' scenario, the expected
incidence ceases to decrease in the first quarter and increases again in
the second quarter. In the ``fast resurgence'' scenario, expected
incidence increases again on a shorter time-scale, typically causing a
large epidemic wave in the second quarter or soon after. Note that for
each scenario, individual trajectories of the base reproduction number
$R_0$ are stochastically simulated and therefore individual incidence curves can substantially differ for a fixed scenario (see SI for details). For each such realization, the model is analyzed with respect to different assumptions of vaccine efficacy and vaccine distribution/administration protocols. This approach permits a direct comparison of different strategies with respect to the same epidemic course in order to clearly evaluate all possible dynamic effects of a change in protocol. As an outcome, we measure the cumulative number of deaths and cumulative number of symptomatic COVID-19 cases in the high-risk group caused within the first 24 weeks (6 months) of the vaccine rollout.
\begin{table*}
    \centering
    \includegraphics[width=0.8\textwidth]{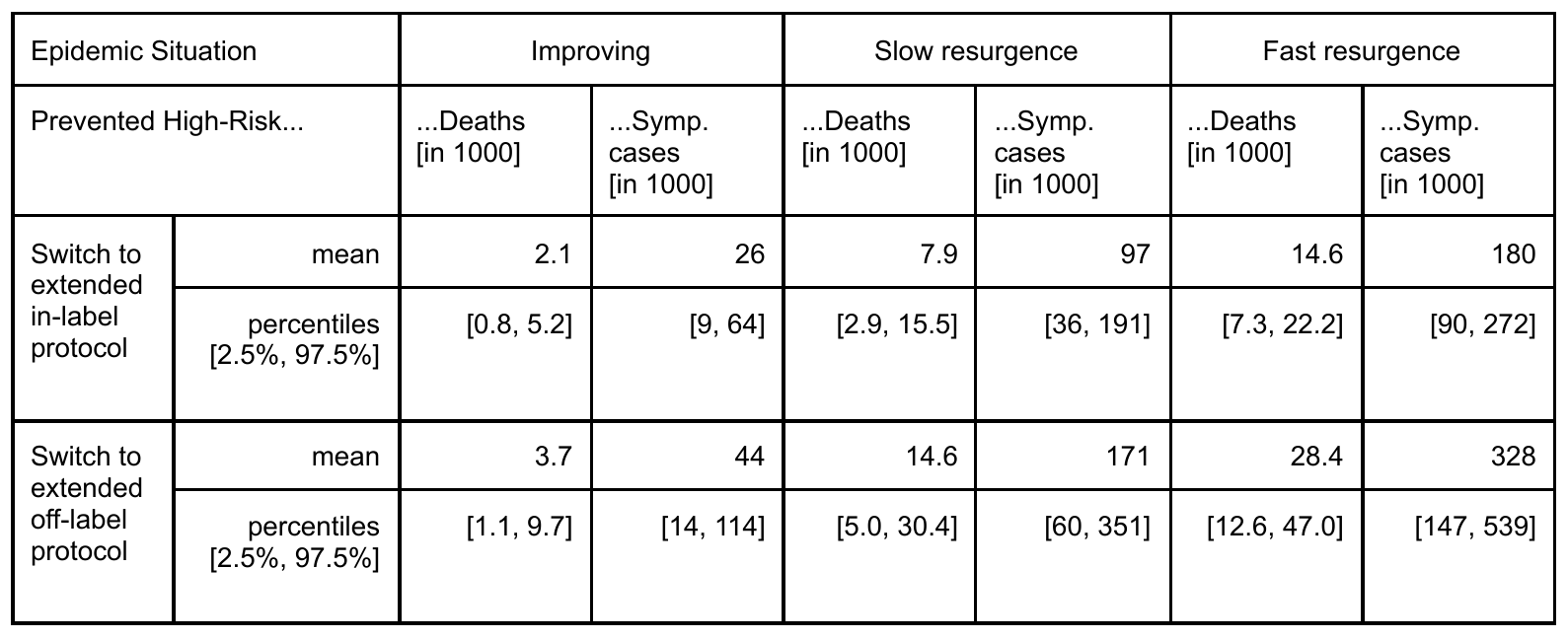}
    \caption{Additionally prevented deaths and prevented symptomatic cases in the high-risk group resulting from a switch to any of the extended vaccination protocols (as compared to the standard protocol), depending on how the epidemic unfolds within the first 6 months of the vaccine rollout. The numbers displayed here represent model results for a situation where in both the first and second quarter, 14 million vaccine doses are available.}
    \label{tab:1}
\end{table*}

\section{Results} \label{sec:results}
\begin{figure*}
    \centering
    \includegraphics[width=\textwidth]{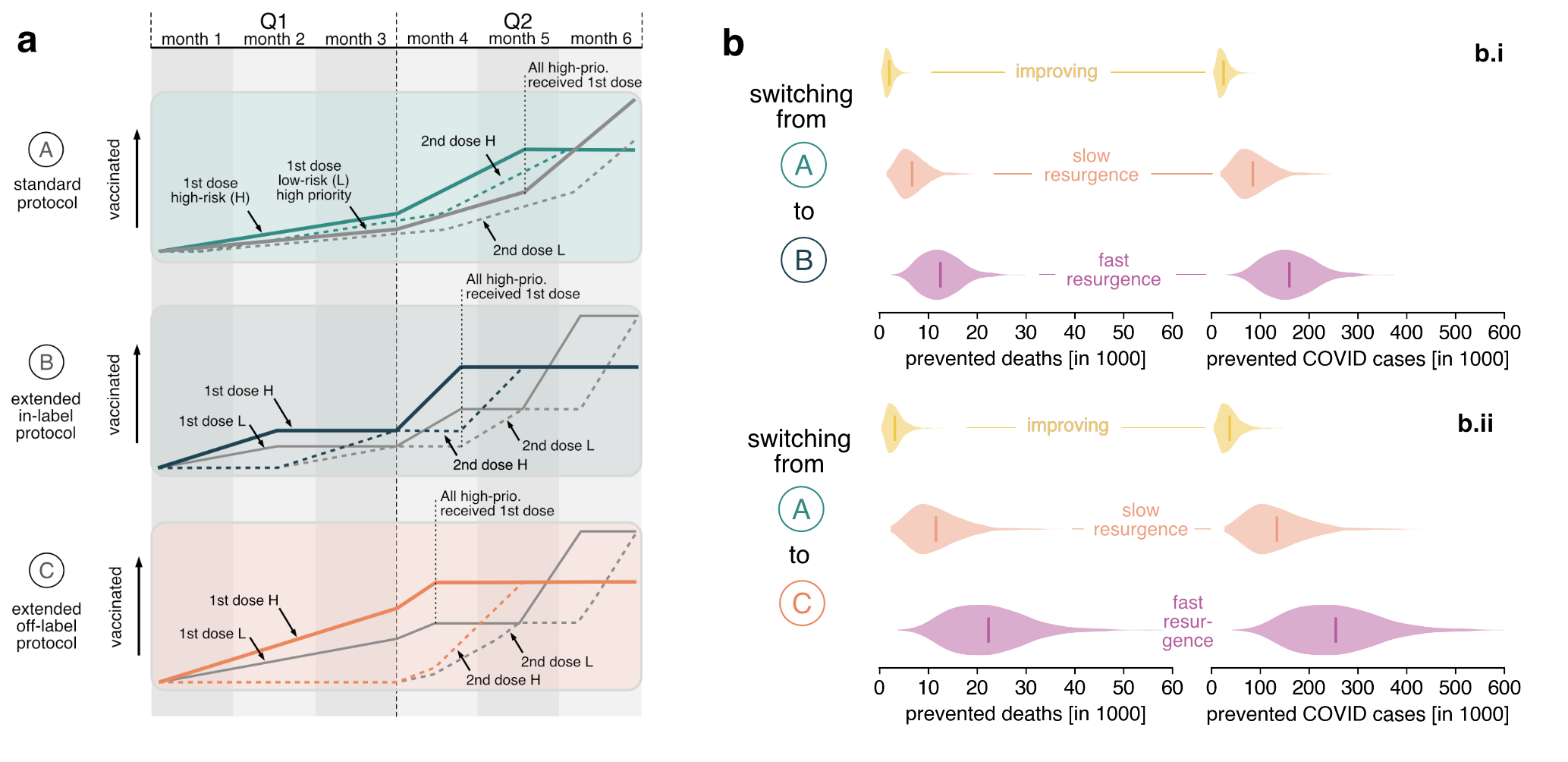}
    \caption{Model results for moderate scarcity of vaccine supply (14 million doses in the first quarter (Q1) and 46 million doses in the second quarter (Q2)). (a) Detailed schematic of vaccine dose distribution. In Q1, doses are distributed in an equal manner to the ``extreme scarcity'' scenario displayed in Fig.~\ref{fig:2}a. In Q2, administering first doses to remaining high-priority individuals is prioritized while ensuring that individuals vaccinated with a first dose will receive a second dose within the defined time frames. After all eligible high-priority individuals have received their first and second doses, the remaining doses are administered to low-priority individuals. Note that in the extended protocols, the time at which all eligible high-priority individuals received their first dose is earlier in the extended protocols as for the baseline protocol. (b) Prevented high-risk group deaths and symptomatic cases when switching from standard to extended protocols (refer to Tab.~\ref{tab:2} for exact numerical values). (b.i) Following the extended in-label protocol instead of the standard protocol will save a number of lives in the four- to low five-digit range. (b.ii) The number of saved lives and prevented symptomatic cases almost doubles when changing to the extended off-label protocol instead.
}
    \label{fig:3}
\end{figure*}

We first investigated the extreme scarcity situation where the number of
delivered doses in Q2 equals the low number of delivered doses in Q1
(see Fig.~\ref{fig:2}). The course of the pandemic in the near future
(Fig.~\ref{fig:2}b) has the most substantial impact on the success of a
change in strategy in terms of the absolute number of high-risk-group
deaths prevented (Fig.~\ref{fig:2}c). If the expected incidence
decreases in the near future, vaccination according to the extended
off-label protocol (Fig.~\ref{fig:2}c.ii) will prevent an average of
3,700 deaths. If the incidence resurges within the first six months of
vaccine roll-out, an average of 14,600 deaths may be prevented by
switching to the extended off-label protocol, reaching a value of 28,400
prevented deaths for a fast resurgence. These values are reduced by 43\%
to 49\% when the vaccination protocol is changed to the extended
in-label protocol instead (see Fig.~\ref{fig:2}c.i and
Tab.~\ref{tab:1}), resulting in 2,100-14,600 deaths prevented. The
advantage of changing to extended protocols decreases by $\leq 25\%$ if
more doses are available in the second quarter (see Fig.~\ref{fig:3} and
Tab.~\ref{tab:2}) but remain of similar order. Overall, we observe a
highly correlated relationship between the intensity of the epidemic and
the success of a change in protocol: the more intense the epidemic, the
more deaths will be prevented by delaying the second dose.

\begin{table*}
    \centering
    \includegraphics[width=0.8\textwidth]{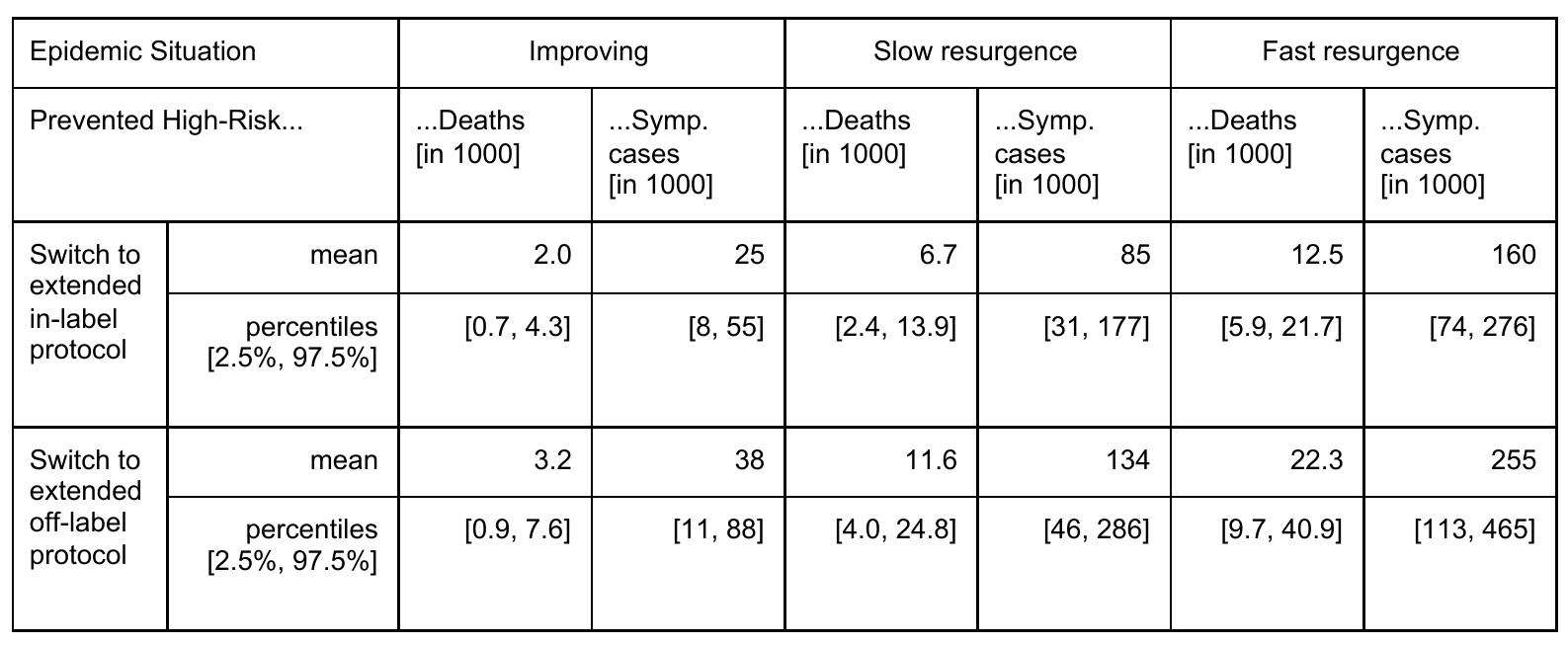}
    \caption{Additionally prevented deaths and prevented symptomatic cases in the high-risk group resulting from a switch to any of the extended vaccination protocols (as compared to the standard protocol), depending on how the epidemic unfolds within the first 6 months of vaccine rollout. The numbers displayed here represent model results for a situation where in the first quarter, 14 million vaccine doses are available, and in the second quarter, 46 million vaccine doses are available.}
    \label{tab:2}
\end{table*}

With respect to the number of symptomatic COVID-19 cases in the high-risk group, a greater reduction can be expected within the first six months of vaccine rollout should any of the extended protocols be implemented (Figs.~\ref{fig:2}c and \ref{fig:3}b), which is caused by the dramatic increase in the number of people protected by the initial vaccine dose in spite of lower efficacy. Note that the number of prevented COVID-19 cases is directly proportional to the fraction of infected individuals that will display symptoms, which is an additional model parameter (here chosen as 83\%). 

In general, the number of prevented symptomatic COVID-19 cases varies between simulations and scenarios. Nevertheless, the model suggests the general rule: the worse the pandemic progresses in the near future, the more worthwhile a change in protocol. However, if the first half of 2021 is followed by another epidemic wave in the second half of the year, a higher number of symptomatic cases in the high-risk group can be expected under the off-label extended protocol compared to the standard protocol, because of assumed reduced vaccine efficacy. However, because protection against death is nearly 100\%, a substantial number of deaths will still be prevented.

A potential drawback of extended protocols is a negative impact on the population’s attitude towards COVID-19 vaccination. We therefore also examined the potential negative effect that a publicly announced change in strategy may have on the confidence of the population and, consequently, the willingness to vaccinate.

\begin{figure*}
    \centering
    \includegraphics[width=\textwidth]{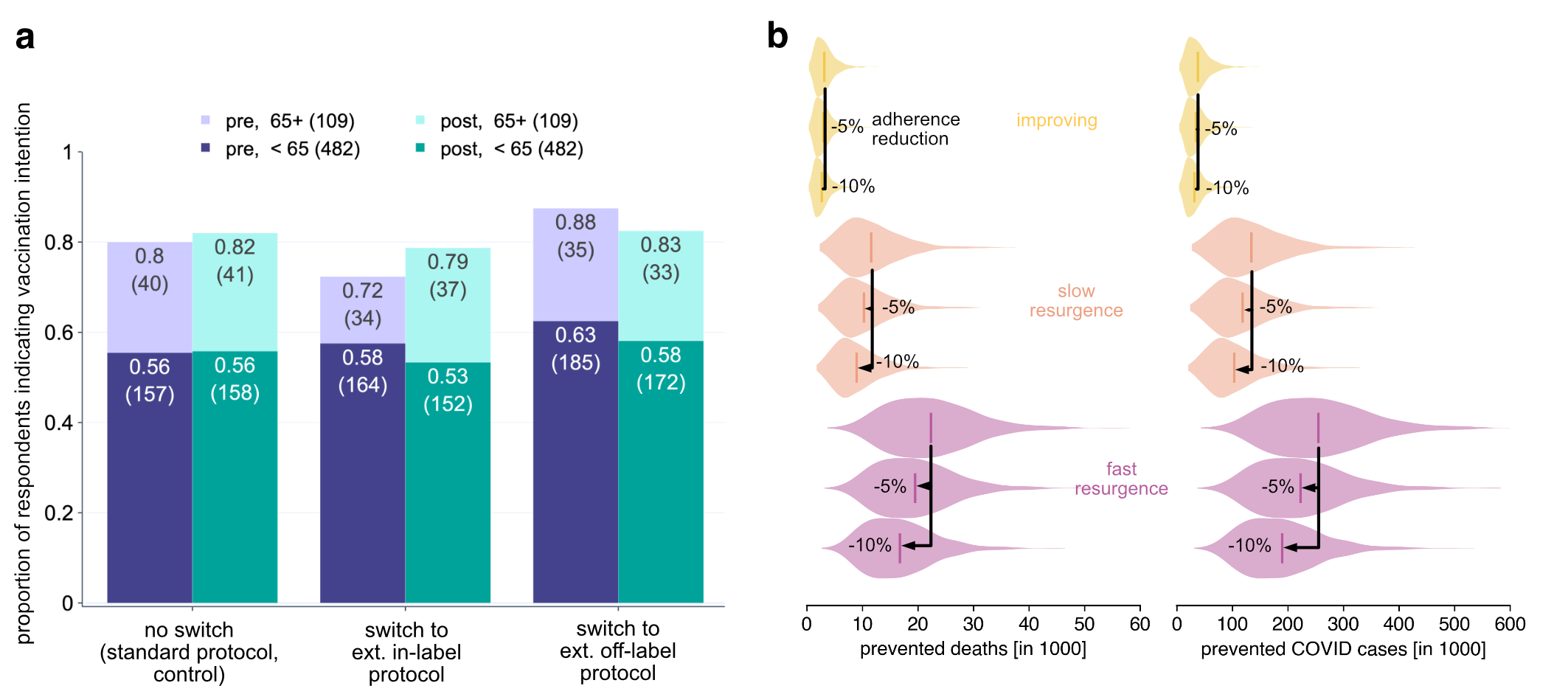}
    \caption{(a) Results of the online survey. Participants’ intention to get vaccinated was collected pre and post manipulation (on a 7 point scale, 1 $=$ not at all vaccinate, 7 $=$ definitely vaccinate; the figure shows the fraction of participants scoring $\geq 5$). They were randomly allocated to three conditions, describing the standard protocol, the extended in-label and extended off-label protocol with preliminary values of their expected consequences (shifting the second dose to 6 vs. 12 weeks after the first dose and preventing 1,000-8,000 or 2,000-13,000 deaths in the next six months, respectively). It was stated that it was still unclear whether the delay of the second dose makes the vaccination less effective overall, but that it is considered unlikely. It was also added whether the delay was in accordance with the licensure. The intention to get vaccinated after learning about the extended protocols did not change substantially and at a maximum of 5\% (participants 65+ in the off-label group). Materials and data at  \cite{betsch_consequences_2021}. (b) Model results assuming that switching protocols to the extended off-label protocol decreases the number of high-risk individuals willing to vaccinate by 5\% and 10\%, respectively. Even with a 10\% reduction in adherence, the number of lives saved and symptomatic cases prevented would still exceed those resulting from a switch to the extended in-label protocol (Fig.~\ref{fig:3}b.i).
}
    \label{fig:4}
\end{figure*}

A survey experiment within the German COVID-19 Snapshot Monitoring (wave
34,
\cite{betsch_monitoring_2020,betsch_germany_2020,betsch_ergebnisse_nodate})
compared the standard protocol to both extended in-label and off-label
strategies ($N = 1001$; methods and data see
ref.~\cite{betsch_consequences_2021}). In a forced-choice setting,
59.9\% opted for the standard protocol, 30.3\% for the extended
in-label, and 8.6\% extended off-label protocol (rest: missing).
However, the decrease in the mean willingness to get vaccinated after
learning about the delay of doses was minimal and non-significant: of
those willing to be vaccinated before the change in strategies, the
average willingness to get vaccinated decreased by 1.5\% in the extended
in-label, and 3.1\% in the extended off-label condition). In the
relevant 65+ age group, the change to the extended off-label protocol
led to a reduction of 5\%. For the extended in-label protocol, we found
a small increase in the number of people willing to get vaccinated
(Fig.~\ref{fig:4}a). However, the sample size in the higher age groups
was rather small, so that the values reported above might be attributed
to noise. Confidence in vaccine safety decreased slightly in the
off-label condition \cite{betsch_beyond_2018}. This change, however, did
not explain the small decrease in the intention to get vaccinated in the
off-label condition. Pre-existing vaccine hesitancy also did not affect
the evaluation of the strategies. In sum, detrimental effects of the
willingness to vaccinate would likely be minimal.

Nevertheless, we compare the number of deaths prevented in the high-risk group by a protocol change, assuming that such a change reduces the number of those willing to vaccinate in the high-risk group by 5\% and 10\% (i.e. from 12 million to 11.4 million and to 10.8 million, respectively), see Fig.~\ref{fig:4}b. We find that even with a 5- to 10-percent reduction in those willing to get vaccinated, there is still a positive impact of a protocol change on the number of deaths as well as prevented symptomatic cases. For a 10\% reduction in the extended off-label protocol, the order of deaths prevented is similar to that of deaths prevented switching to the extended in-label protocol with a 0\% reduction. Approximately, a 30\% reduction in vaccination willingness is required to observe negative effects of the extended off-label protocol compared to the standard protocol regarding the number of prevented COVID-19 cases in the high-risk group (see SI).

If we assume that vaccine efficacy against disease after the first dose is only 80\% and that this value is maintained, but not increased, by administering the second dose after 12 weeks (which corresponds to a 100\% increase in the assumed probability of showing symptoms after infection), the absolute number of symptomatic cases prevented by the change in strategy decreases by approximately 26-42\%. Even if we increase the post-vaccination IFR drastically by a factor of 50 (equivalent to reducing efficacy against death to 95\%), the number of prevented deaths decreases only by 11--18\% (see SI).

Conservatively, and because exact values are unknown at the time of writing, we assumed a zero transmission reduction for vaccinated, yet infected individuals. If we assumed large transmission reductions (90\% for both doses), the number of prevented deaths and prevented symptomatic cases would increase by a low percentage for both extended in-label and off-label protocols. If, instead, we analyzed an extreme and unbalanced scenario where the first dose does not reduce transmission probability for such individuals but the second dose reduces transmission by 90\%, a switch to extended protocols would still be beneficial, but would reduce the prevented deaths and cases by 13--30\% as compared to results obtained using the original assumptions.

These analyses suggest that earlier protection of a larger group of high-risk individuals has a greater systemic protective effect in a hypothetical off-label vaccination protocol, even if switching protocols might lead to a reduction in vaccine adherence or vaccine efficacy.

\section{Discussion and conclusion} \label{sec:discussionconclusion}

The results suggest that switching to an alternative vaccination strategy (12-week delay of the second dose) of the currently available mRNA vaccines against COVID-19 could prevent an additional number of deaths in the four to five digit range for hypothetical vaccine rollout scenarios in Germany. Such benefits are reached should incidence resurge, for instance in the case of the spread of one of the recently discovered more contagious viral variants---given that the existing vaccines are effective against these variants. This is true even for a low reduction in willingness to vaccinate induced by a change in protocol. A switch to extended protocols is even more beneficial when supply of vaccines remains limited. If nonpharmaceutical interventions show effect in the first half of 2021 and the incidence remains low over the first half of 2021, however, the number of deaths prevented by a change in protocol drops to lower values in the three- to four-digit range. The respective values are likewise reduced if vaccine efficacy is lower than anticipated but remain of similar magnitude nonetheless. If a change in strategy were adopted, damage to vaccine confidence and willingness to get vaccinated needs to be prevented. Results of the online survey in Germany showed that protocol changes would likely have only minimal detrimental effects on the willingness to get vaccinated. However, a shift in protocols would increase the need for communication measures as the survey also showed that deviating from the standard protocol was not the best preferred solution despite a high number of deaths prevented. In the course of a feasibility study it could be determined, for example, to what extent a voluntary three-month waiver of the second dose allows additional vaccinations without reducing the willingness to vaccinate, and whether the higher vaccination rate achieved in this way justifies potential other risks.

Risks related to delaying the second dose of vaccination which are not considered here include (i) the possibility of infection enhancing antibodies (antibody dependent enhancement, ADE), (ii) the increased likelihood of the development of escape mutations of the virus, and (iii) that the high efficacy of the first dose does not persist. 

ADE has not been shown after vaccination so far, neither after first nor second dose. Given the large number of people immunized so far and that ADE is likely to occur early in the course of vaccination, the risk appears to be small. An increased mutation pressure through the combination of imperfect immunity and high case loads (peak risk) cannot be ruled out. However, it is unclear if the delay of the second dose increases or decreases mutation pressure, since an early first dose reduces case load but may at the same time increase selection pressure, the net effect being unclear. Loss of protection of neutralizing antibody titers between first and second dose in extended second dose protocols is possible but unlikely given high titers following first dose mRNA vaccination.

Despite the benefits and advantages of changing the standard protocol and delaying the second dose that persist even if vaccination adherence should drop in response to changing protocols, the results must be viewed as part of a wider context and must be carefully weighed against a number of risks that are difficult to quantify and beyond the scope of the current analysis. In fact, the model does not account for several of the widely discussed and difficult-to-calculate risks of postponing the second vaccination that must be taken into account when viewed as part of a practical public health guide line. The current model is limited and intentionally designed to estimate the epidemiological consequences in terms of the expected change in the number of deaths and COVID-19 cases only, and helps with the assessment of how benefits or disadvantages may depend on parameters of the system.

\section*{Acknowledgments}
\label{sec:acknowledgments}
BFM is financially supported as an \emph{Add-On Fellow for Interdisciplinary Life Science} by the Joachim Herz Stiftung. MMH was supported by the European Union’s Horizon 2020 research and innovation programme under grant agreement No 101003480 and by German Federal Ministry of Education and Research within the project CoViDec (FKZ: 01KI20102). We express our gratitude to C.~Drosten for helpful comments during the research process.  BFM would like to thank W.~Wu for productive discussions. DB would like to thank I.~Mortimer and V.~Hardapple for valuable comments on the manuscript.

\let\oldUrl\url
\renewcommand{\url}[1]{\href{#1}{embedded in PDF}}
\bibliographystyle{unsrtnat}
\bibliography{references}

\appendix
\section*{Supplementary Information}
\UseRawInputEncoding 

\section{Model}

We first define a deterministic compartmental infection model based on a susceptible-exposed-infected-recovered
(SEIR) model with additional compartments for vaccination status and
parallel counting of symptomatic cases, confirmed cases, and fatalities.
We further consider the population to be structured according to risk-group
affiliation. Such models are well-established and -studied \cite{keeling_modeling_2011,anderson_infectious_2010}.

In this model, individuals are part of either low-risk ($Y$) or high-risk
($A$) groups of sizes $N^{Y}=\mathrm{const}.$ and $N^{A}=\mathrm{const.}$,
respectively, such that the total population size is given as $\mathcal{N}=N^{Y}+N^{A}$.
An affiliation with group $a$ will be denoted by a superscript $(\cdot)^{a}$
in the following but will rarely be used explicitly because individuals
cannot change status between these population groups by any means.

Initially, all individuals of group $a$ are ``susceptible'' with
protection level zero (received zero vaccine doses), i.e.~are part
of compartment $S_{0}^{a}\equiv S_{0}$ (note that we also denote
the probability of any person of group $a$ to be susceptible as $0\leq S_{0}^{a}\leq1$
and similarly for other compartments). Individuals may be vaccinated
according to three protocols. In the standard protocol, people are
vaccinated with a single dose, while the second dose is given out
immediately but held back for approximately three weeks to be given
to the same person. This dose therefore cannot be used on other individuals
during this time. In the ``extended in-label protocol'', a person
is vaccinated with a single dose, but the second dose will not be
held back. Instead, the remaining dose is administered to a second
person. After $42$ days, a second dose will be given to both
individuals, which represents the maximal time between the administration
of two doses by the BNT162b2 vaccine for the vaccination to
be considered ``in-label'', i.e. administered as recommended by
the manufacturer. The third, ``extended off-label protocol'' postpones
the second dose for 12 weeks, i.e.~$84$ days, which is
considered an ``off-label'' vaccination but allows one to give twice as many individuals a first protection in the
first quarter when vaccine doses are scarce. Note that throughout this paper, we define a quarter to have a duration of 84 days and 6 months to comprise 168 days.

\begin{figure*}
\begin{centering}
\includegraphics[width=0.75\textwidth]{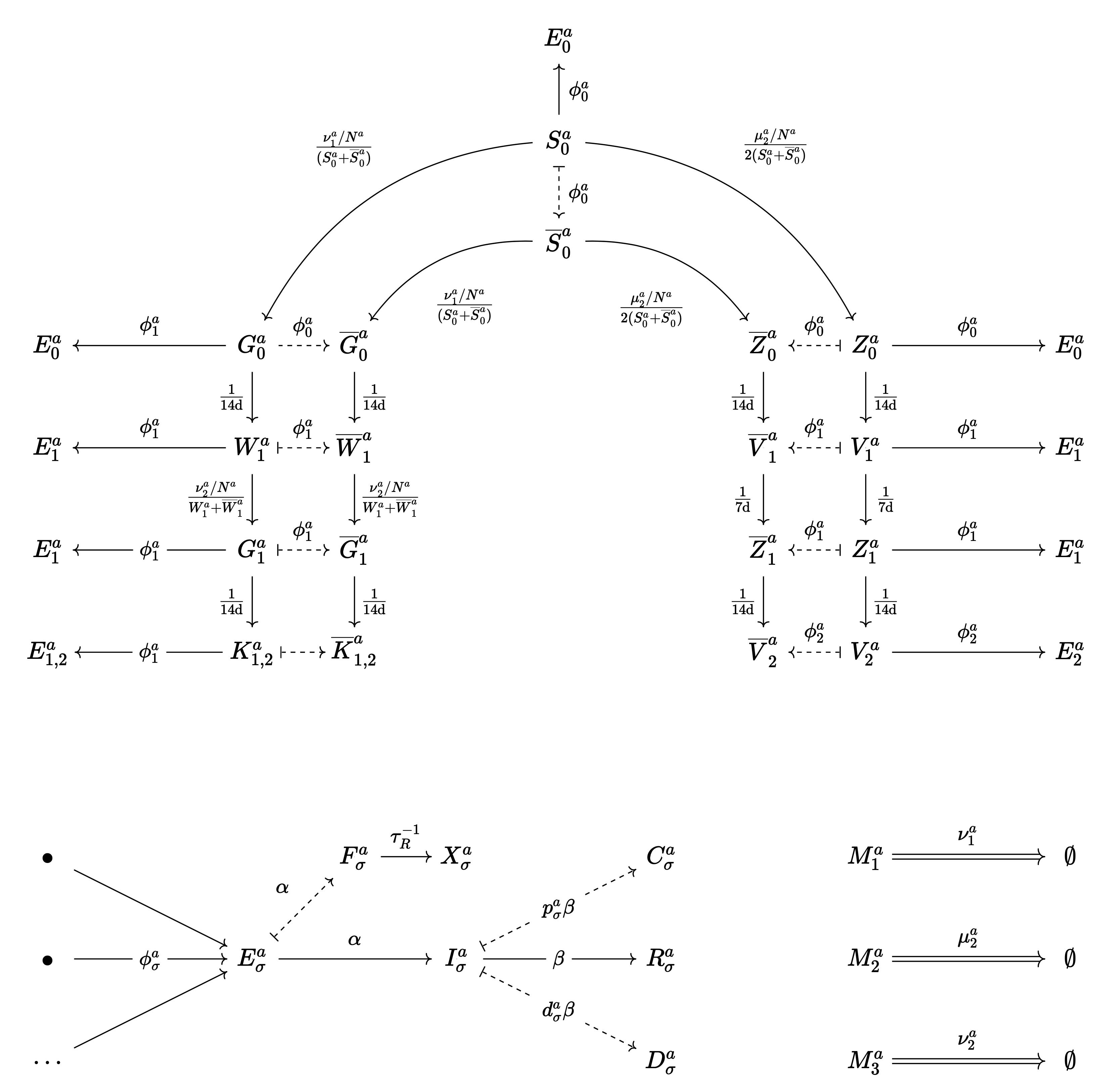}\caption{Schematic figure of the parallel infection/vaccination dynamics. Initially, an individual is fully susceptible
to infection ($S_{0}^{a}$). A person might become infected with rate
$\phi_{0}^{a}$, at which point it transitions to the infection dynamics
that are run in parallel to the vaccination dynamics. Parallel counting is introduced here
because in Germany, infection status is not checked when administering
vaccines, i.~e.~vaccines will be distributed equally to people that
already went through the infection process. In order to simplify notation, we count infected individuals double, once to be regarded in the infection process and once to be regarded for the vaccination process. Individuals that received
the first dose obtain a first protection level about 14 days after.
Following the standard protocol (doses depleted from $M_{2}^{a}$, right arm of the schema), first-level protected individuals will
receive the second, reserved dose 7 days after and reach the second
protection level after an additional 14 days. Note that while $M_{2}^{a}$
is linearly depleted with rate $\mu_{2}^{a}$, the number of corresponding
individuals with first-level protection increases only with rate $\mu_{2}^{a}/2$
which reflects that per first-dose individual, two doses are depleted from the available vaccine stock.
In the extended protocols (left arm of the schema), the second dose
is administered according to a phase- and risk-specific reservoir
for second doses ($M_{3}^{a}$). Depending on phase-specific definitions,
individuals receive a first- or second-level protection by the second
dose. An individual's infection status is determined by its risk group
affiliation $a$ and protection level $\sigma$. For each $\sigma$ and $a$, susceptibility
is reduced by a specific value $r_{\sigma}^{a}$. If infected, an
individual first reaches the ``exposed'' status $E_{\sigma}^{a}$.
From there, it progresses to the ``infectious'' status $I_{\sigma}^{a}$
after an average duration of $1/\alpha$, from which they will be
``removed'' after an average duration of $1/\beta$. A fraction
$p_{\sigma}^{a}$ of removed individuals will have had symptoms (classified
as $C_{\sigma}^{a}$) and a fraction $d_{\sigma}^{a}$ of removed
individuals will have had fatal outcomes. In parallel, individuals
that become infectious will be counted after an average reporting
delay of $\tau_{R}$ by entering the $X_{\sigma}^{a}$ compartment.}
\par\end{centering}
\end{figure*}

Model equations will be defined in separate but consecutive phases
that are defined for time intervals $[t_{0},t_{0}+T)$. The total
amount of remaining vaccine doses at time $t$ will be denoted as
$M_{1}^{a}(t)$ for the first shot in the extended protocols, $M_{2}^{a}(t)$
for all doses administered according to the standard protocol, and $M_{3}^{a}(t)$
denoting the number of doses reserved for second shots in the extended
protocols. We assume a constant vaccine supply rate and therefore
a linear depletion of both vaccine doses and susceptibles with rates
\begin{align*}
\nu_{1}^{a} & =\frac{n_{1}^{a}}{T}\\
\mu_{2}^{a} & =\frac{n_{2}^{a}}{T}\\
\nu_{2}^{a} & =\frac{n_{3}^{a}}{T}
\end{align*}
such that
\begin{align*}
\frac{d}{dt}M_{1}^{a} & =-\nu_{1}^{a}\\
\frac{d}{dt}M_{2}^{a} & =-\mu_{2}^{a}\\
\frac{d}{dt}M_{3}^{a} & =-\nu_{2}^{a}
\end{align*}
where 
\begin{align*}
M_{1}^{a}(t_{0}) & =n_{1}^{a}\\
M_{2}^{a}(t_{0}) & =n_{2}^{a}\\
M_{3}^{a}(t_{0}) & =n_{3}^{a}.
\end{align*}
Here, $n_{i}^{a}$ refers to the total number of vaccine doses available
to population group $a$ following shot $i$.
Upon receiving a vaccine dose, a susceptible individual transitions
to either $G_{0}$ (extended protocols) or $Z_{0}$ (standard protocol).
In many countries, the employed vaccine roll-out strategies do not
require individuals to prove that they have not yet been infected
with SARS-CoV-2. Therefore, we consider two parallel vaccine rollouts
and divide the population in ``never had a SARS-CoV-2 infection''
(denoted by default compartment symbols such as $S_{0}$) and ``currently
are infected with SARS-CoV-2 or have been infected in the past''
(denoted by overlined compartment symbols $\overline{S}_{0}$). When
picking a non-vaccinated individual to be vaccinated at random, the probability
to pick a truly susceptible person is $S_{0}/(S_{0}+\overline{S}_{0})$,
which means that the vaccine-related depletion of individuals with
vaccine status zero is given as
\begin{align*}
\frac{d}{dt}S_{0} & =-\left(\frac{\nu_{1}}{N}+\frac{\mu_{2}}{2N}\right)\frac{S_{0}}{S_{0}+\overline{S}_{0}}-\phi_{0}S_{0}.
\end{align*}
Note that following the standard protocol, only half the number of
people can be vaccinated per day initially as compared to the extended
protocol (as explained above), hence a factor $1/2$ is introduced.
The transitional term $\phi_{0}S_{0}$ accounts for infections that
happen in parallel to the vaccination process. We will define the
force of infection $\phi_{\sigma}^{a}\equiv\phi_{\sigma}$ further
below. Note that the depletion of already infected individuals is
given as
\[
\frac{d}{dt}\overline{S}_{0}=-\left(\frac{\nu_{1}}{N}+\frac{\mu_{2}}{2N}\right)\frac{\overline{S}_{0}}{S_{0}+\overline{S}_{0}}+\phi_{0}S_{0}.
\]
In the following we will omit the explicit definitions of changes
in infectious/infected counterparts $\overline{(\cdot)}$, as their
equations of motion are equal to their respective non-infected counterparts
bar an opposite sign in the infection terms.

Individuals that received the first dose (i.e.~ are part of $Z_{0}$,
$G_{0}$, or their respective overlined counterparts) are reaching
protection with protection level 1 after an average duration of
$\tau_{M}=14\mathrm{d}$ such that
\begin{align*}
\frac{d}{dt}G_{0} & =\frac{\nu_{1}}{N}\frac{S_{0}}{S_{0}+\overline{S_{0}}}-\frac{1}{\tau_{M}}G_{0}-\phi_{0}G_{0},\\
\frac{d}{dt}Z_{0} & =\frac{\mu_{2}}{2N}\frac{S_{0}}{S_{0}+\overline{S_{0}}}-\frac{1}{\tau_{M}}Z_{0}-\phi_{0}Z_{0}.
\end{align*}
A person that was vaccinated according to the standard protocol will
receive the second dose approximately $\tau_{D}=21\mathrm{d}-\tau_{M}=7\mathrm{d}$
after reaching the first protection level. Afterwards, it takes
another $\tau_{M}$ days to reach protection level 2. The remaining
equations of motion for all standard-protocol individuals therefore
read

\begin{align*}
\frac{d}{dt}V_{1} & =\frac{1}{\tau_{M}}Z_{0}-\frac{1}{\tau_{D}}V_{1}-\phi_{1}V_{1}\\
\frac{d}{dt}Z_{1} & =-\frac{1}{\tau_{M}}Z_{1}+\frac{1}{\tau_{D}}V_{1}-\phi_{1}Z_{1}\\
\frac{d}{dt}V_{2} & =\frac{1}{\tau_{M}}Z_{1}-\phi_{2}V_{2}.
\end{align*}
Regarding the extended protocols, individuals that received the first
dose and obtained the first protection level remain in compartment
$W_{1}$ until they are vaccinated again with rate $\nu_{2}$ in a
consecutive phase. The equations of motion read
\begin{align*}
\frac{d}{dt}W_{1} & =\frac{1}{\tau_{M}}G_{0}-\frac{\nu_{2}}{N}\frac{W_{1}}{W_{1}+\overline{W}_{1}}-\phi_{1}W_{1},\\
\frac{d}{dt}G_{1} & =\frac{\nu_{2}}{N}\frac{W_{1}}{W_{1}+\overline{W}_{1}}-\phi_{1}G_{1}-\frac{1}{\tau_{M}}G_{1}\\
\frac{d}{dt}K_{1} & =\begin{cases}
\frac{1}{\tau_{M}}G_{1}-\phi_{1}K_{1} & \mathrm{>6\ weeks\ postponed},\\
-\phi_{1}K_{1} & \mathrm{\leq6\ weeks\ postponed},
\end{cases}\\
\frac{d}{dt}K_{2} & =\begin{cases}
-\phi_{2}K_{2} & \mathrm{>6\ weeks\ postponed},\\
\frac{1}{\tau_{M}}G_{1}-\phi_{2}K_{2} & \mathrm{\leq6\ weeks\ postponed.}
\end{cases}
\end{align*}
Even though individuals of compartment $W_{1}$ receive a second dose
after 12 weeks, we will conservatively assume that postponing the
second dose merely upholds the protection achieved by the first dose
and does not enhance it when administered after a duration longer
than 6 weeks (as such a procedure is an off-label vaccination). For
vaccinations by the in-label extended protocol, we instead assume
that the second protection level is reached.

In total, we have
\[
\frac{d}{dt}\Big(S_{0}+G_{0}+Z_{0}+G_{1}+W_{1}+Z_{1}+K_{1}+V_{1}+V_{2}+K_{2}+\mathrm{i.c.}\Big)=0,
\]
such that the total number of people in each population group remains
constant (here, ``i.c.'' refers to ``infectious counterparts'',
i.e. the respective overlined compartments). Note that we define the sum in the equation above to be equal to 1.

In parallel to the vaccination process, individuals can become infected
and infectious. We categorize susceptible individuals of group $a$
and protection level $\sigma\in\left\{ 0,1,2\right\} $ as 
\begin{align*}
\mathcal{S}_{0}^{a} & =\left\{ S_{0}^{a},G_{0}^{a},Z_{0}^{a}\right\} \\
\mathcal{S}_{1}^{a} & =\left\{ G_{1}^{a},Z_{1}^{a},W_{1}^{a},V_{1}^{a},K_{1}^{a}\right\} \\
\mathcal{S}_{2}^{a} & =\left\{ V_{2}^{a},K_{2}^{a}\right\} .
\end{align*}
Then, for each group $a$ and protection level $\sigma$, the infection
process is defined by the equations
\begin{align*}
\frac{d}{dt}E & =\phi\sum_{H\in\mathcal{S}}H-\alpha E\\
\frac{d}{dt}I & =\alpha E-\beta I\\
\frac{d}{dt}R & =\beta I\\
\frac{d}{dt}D & =d\beta I\\
\frac{d}{dt}C & =p\beta I.
\end{align*}
Here, all quantities except $t$, $\alpha$, and $\beta$ should be
read as explicitly depending on group affiliation $a$ and protection level $\sigma$ (sub- and superscripts are omitted for readability).
We assume a universal average latency time of $\alpha^{-1}=3.2\mathrm{d}$
after which exposed individuals $E$ transition to compartment $I$.
We further assume a universal infectious period of $\beta^{-1}=6.7\mathrm{d}$
after which individuals become non-infectious and immune or otherwise
removed from the process ($R$). We count individuals who pass away
in compartment $D$ as the proportion $d_{\sigma}^{a}\equiv d$ of
$R$. Additionally, we track the cumulative amount of people who will
become symptomatic (show symptoms of COVID-19) in compartment $C$
as the ratio $p_{\sigma}^{a}\equiv p$ of $R$. Note that
\[
\frac{d}{dt}\sum_{\sigma=0}^{2}\Big(E_{\sigma}^{a}+I_{\sigma}^{a}+R_{\sigma}^{a}+\sum_{H\in\mathcal{S}_{\sigma}^{a}}H\Big)=0,
\]
i.e.~the total population size in group $a$ remains constant at all
times (we require the sum in the equation above to be equal to unity).
Since infected individuals are counted twice in separate parts of
the model, we have
\[
\frac{d}{dt}\sum_{\sigma=0}^{2}\Big(\sum_{\bar{H}\in\bar{\mathcal{S}}_{\sigma}^{a}}\bar{H}+\sum_{H\in\mathcal{S}_{\sigma}^{a}}H\Big)=0,
\]
as well. The total force of infection on any susceptible individual
$H_{\sigma}^{a}\in(\mathcal{S}_{0}^{a}+\mathcal{S}_{1}^{a}+\mathcal{S}_{2}^{a})$
is given as 
\begin{equation}
\phi_{\sigma}^{a}(t)=\frac{\mathcal{R}^{YY}(t)\beta}{C^{YY}}(1-r_{\sigma}^{a})\sum_{a'\in\{Y,A\}}\sum_{\sigma'=1}^{2}(1-k_{\sigma'}^{a'})C^{aa'}I_{\sigma'}^{a'}(t).\label{eq:force_of_infection}
\end{equation}
Here, $C^{aa'}$ is the average number of contacts a person of group
$a$ has with a person of group $a'$ (temporal and ensemble average).
We scale the force of infection globally by the in-group temporal
reproduction number $\mathcal{R}^{YY}(t)$ of the low-risk group.
We further decrease an individual's susceptibility based on group
affiliation $a$ and protection level $\sigma$ using an assumed
susceptibility reduction $r_{\sigma}^{a}$. Similarly, the infectivity
of an infectious person of group $a'$ and protection level $\sigma'$
is reduced based on a transmissibility reduction $k_{\sigma'}^{a'}$.

In addition to the default infection process, we count the cumulative
number of all infected people $X$ with reporting delay $\tau_{R}$
as 
\begin{align*}
\frac{d}{dt}F & =\alpha E-\frac{1}{\tau_{R}}F\\
\frac{d}{dt}X & =\frac{1}{\tau_{R}}F.
\end{align*}
The full set of equations is given in Eqs.~(\ref{eq:model_ODEs_begin}-\ref{eq:model_ODEs_end}).

In addition to the model equations, we find the daily incidence of
cases in age group $a$ to be
\begin{equation}
J^{a}=N^{a}\sum_{\sigma=0}^{2}\frac{d}{dt}X_{\sigma}^{a}=\frac{N^{a}}{\tau_{R}}\sum_{\sigma=0}^{2}F_{\sigma}^{a}.
\label{eq:incidence_age_group}
\end{equation}
As argued below, we assume that only a fraction $\delta^{a}$ of all
infected individuals of risk group $a$ are reported. This implies
that the total reported daily incidence is given as
\begin{equation}
J=\delta^{A}J^{A}+\delta^{Y}J^{Y}.\label{eq:incidence_total}
\end{equation}
Furthermore, following the derivation in \cite{heng_approximately_2020,wallinga_how_2007}
for SEIR models, we define the temporal effective reproduction number
as 
\[
\mathcal{R}_{\mathrm{eff}}=(1+\Lambda/\alpha)(1+\Lambda/\beta)
\]
where we evaluate the growth rate $\Lambda$ as based on the incidence
as 
\[
\Lambda=\frac{d\mathrm{log}J}{dt}.
\]

In this study, we are interested in how different vaccine distribution protocols influence the cumulative number of deaths in the high-risk group and the cumulative number of symptomatic cases in the high-risk group. Hence, we evaluate the observables
\begin{align*}
    \Delta D^A &= N^A\sum_{\sigma=0}^2\Big[D_{\sigma,\mathrm{standard}}^A(168\mathrm d) - D_{\sigma,\mathrm{extended}}^A(168\mathrm d)\Big]\\
    \Delta C^A &= N^A\sum_{\sigma=0}^2\Big[C_{\sigma,\mathrm{standard}}^A(168\mathrm d) - C_{\sigma,\mathrm{extended}}^A(168\mathrm d)\Big].
\end{align*}
to compare the impact of any of the extended protocols.
Here, $(\cdot)^A_{\sigma,\mathrm{protocol}}(168\mathrm d)$ represents the value of compartment $(\cdot)$ at the end of the respective 6-month vaccine distribution protocol, i.~e.~their respective cumulative count.
\section{Population structure and parameter choices}
The German national vaccine rollout strategy entails a tier-based
provision of vaccines to people of different risk groups \cite{vygen-bonnet_beschluss_2020}.
Here, we consider the first three tiers that comprise most of the
individuals at high risk regarding death by COVID-19. Tier 1 contains
individuals of age >80 and inhabitants of nursing homes ($\approx6.4$
million) as well as high-risk-of-exposure healthcare workers. Tier 2 entails
individuals aged 75\textendash 79, individuals with diagnosed dementia,
people with mental disabilities (all mentioned in total $\approx5.7$
million), and other health care workers. Tier 3 contains, besides others,
individuals aged 70\textendash 74 ($\approx3.6$ million). In total, the high-risk
group contains about
15.7 million people. This group will be referred to by the superscript
$A$ in the following. Tier 1 and tier 2 alone total about $>15.6$ million
individuals of which 12.1 million are in the high-risk group (the rest being medical workers).

In order to structure the population in a way that reflects both the
size of the high risk group as well as the higher age of people within
the risk group, we choose 
\begin{align*}
C^{YY} &= 7.781659\\
C^{YA} &= 0.65415356\\
C^{AA} &= 1.6846154\\
C^{AY} &= 2.821114
\end{align*}
with $N^{A}=15,293,178$, and $N^{Y}=65,953,623$, which is the contact
structure one obtains when distributing the population into age strata
$[0,65)$ and $65+$ using data from the POLYMOD study with the open
source R software ``socialmixr'' \cite{mossong_social_2008,funk_socialmixr_2020}.
We find that the contact matrix does not change substantially when
choosing age strata {[}0,70) and 70+ which is closer to our high-risk
group criteria but yields a much smaller high-risk group population
size (due to the data being from 2005).

We calibrate the infection fatality ratio (IFR) for both risk groups
based on the course of the pandemic in Germany in 2020 using data
for age strata {[}0,59) (an approximation to group $Y$) and 60+ (approximation
to group $A$) due to the lack of more fine-grained data and find
that by choosing IFRs of
\begin{align*}
d_{0}^{Y} & =0.028\%\\
d_{0}^{A} & =6.250\%
\end{align*}
the progression of the pandemic in Germany in 2020 is reflected reasonably
well by the model (see Sec. \ref{sec:Calibration-Simulation}). We calibrate the model by comparing the 7-day running average of daily new reported cases and daily new deaths of official case counts \cite{robert_koch_institute_fallzahlen_nodate}. These IFR
values are within the reported bounds of a meta-review study
on IFR by age \cite{levin_assessing_2020} and correspond to a population-wide IFR of 1.15\%. The exact value of the IFR is, however, of less importance for our analysis. In order to accurately model the situation, we have to calibrate the model case fatality rate (CFR) to be equal to the reported CFR. Varying the IFR will have to be compensated by varying the respective dark factors accordingly to match the empirical CFR. If, then, reported incidence curves remain constant, the number of deaths will approximately remain constant, as well.

Based on the phase 3 studies for the Biontech/Pfizer and Moderna vaccines,
we assume that the IFR decreases to diminishingly small values beginning 14 days after
receiving one or two doses. In neither study, any participant that
received the vaccine passed away by COVID-19 and only one person developed
a severe case of COVID-19 as compared to 4 severe cases in the placebo
group \cite{vygen-bonnet_beschluss_2020}. We therefore assume
vaccine efficacies of 
\begin{align*}
e_{\mathrm{death,}1}^{a} & =99.9\%,\\
e_{\mathrm{death,}2}^{a} & =99.99\%,
\end{align*}
and set
\begin{align*}
d_{1}^{a} & =d_{0}^{a}\times(1-e_{\mathrm{death,}1}^{a})/(1-r_1^a)\\
d_{2}^{a} & =d_{0}^{a}\times(1-e_{\mathrm{death,}2}^{a})/(1-r_2^a).
\end{align*}

We assume a uniform transmissibility reduction of $k_{\sigma}^{a}=0\%$
for all groups and protection levels. Vaccine studies showed a
reduction in the rate of viral shedding following SARS-CoV-2 infection
in various animal models, yet vaccinated humans will manifest few
if any symptoms\textemdash consequently, these individuals would not
change their behavior (a typical cause for an effective transmission
reduction in symptomatic cases), which we assume results in a diminishing
net transmission reduction.

Regarding a susceptibility reduction of vaccinated individuals, preliminary
results from Israel suggest a 51\% efficacy against infection with
SARS-CoV-2 12 days after receiving the first dose \cite{chodcik_effectiveness_2021}
which we assume translates to a $\approx50$\% reduction in susceptibility
for those with protection level 1, i.e. $r_{1}^{a}=0.5$. We further
assume that this protection increases to $r_{2}^{a}=0.6$ for the
second protection level.

While irrelevant for the dynamics of the system, we aim to estimate
the number of prevented symptomatic COVID-19 cases, as well. Following
ref.~\cite{byambasuren_estimating_2020}, we set a population-wide
probability for symptomatic infection of
\[
p_{0}^{a}=0.83.
\]
We further implement a vaccine efficacy 14 days after receiving a second dose
in-label as 
\[
e_{\mathrm{COVID,}2}^{a}=0.95.
\]
Data from the BNT162b2 phase 3 study further suggests an efficacy
of 
\[
e_{\mathrm{COVID},1}^{a}=0.9
\]
12-14 days after receiving the first dose \cite{polack_safety_2020}.
We implement these values as 
\begin{align*}
p_{1}^{a} & =p_{0}^{a}\times(1-e_{\mathrm{COVID},1}^{a})/(1-r_1^a),\\
p_{2}^{a} & =p_{0}^{a}\times(1-e_{\mathrm{COVID},2}^{a})/(1-r_2^a).
\end{align*}

\section{Calibration simulation}

\label{sec:Calibration-Simulation}

\begin{figure*}
\begin{centering}
\includegraphics[width=1\textwidth]{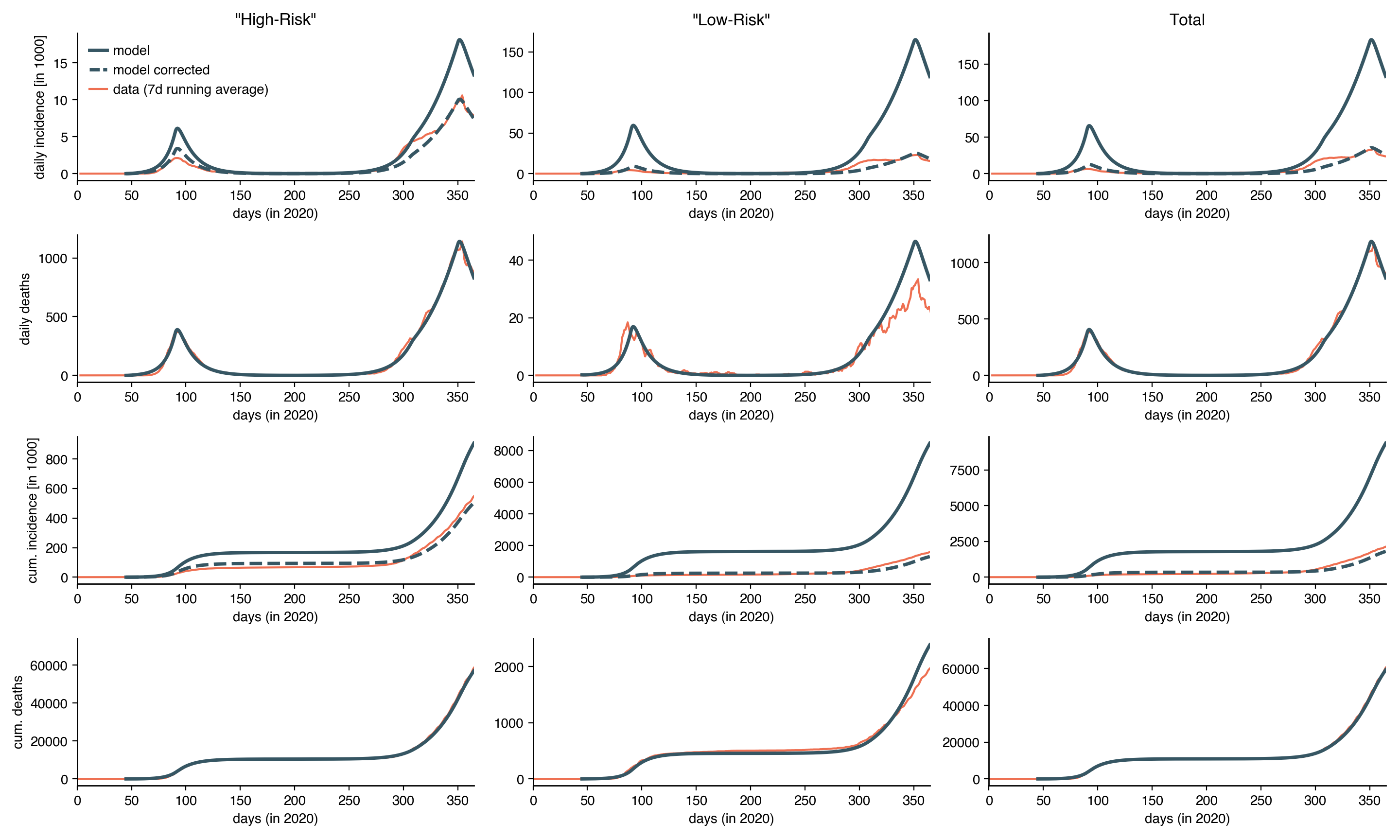}
\caption{\label{fig:Calibration-simulation}Calibration simulation. We initiate
the model as described in Sec.~\ref{sec:Calibration-Simulation}
and use the temporally varying low-risk temporal basic reproduction
number as given in Eq.~(\ref{eq:R_calib}) to scale the force of
infection as defined in Eq.~(\ref{eq:force_of_infection}). The basic
reproduction number Eq.~(\ref{eq:R_calib}) was chosen such that
daily high-risk group model fatalities match the 7-day running average
of daily fatalities in the 60+ age group in Germany during 2020. Model
incidence (thick blue curves) overestimates the empirical incidence
by a factor of $1/\delta^{A}=1.8$ and $1/\delta^{Y}=6.5$, respectively,
which is in line with reported dark factors obtained via sero-prevalence
studies in Germany \cite{streeck_infection_2020,santos-hovener_seroepidemiologische_2020}.
In order to closely match fatalities in the low-risk group in the
model to the age group of $<60$ in the data, we chose an infection
fatality rate of $d_{0}^{Y}=0.028\%$.}
\par\end{centering}
\end{figure*}

As initial conditions we choose $I_{0}^{Y}(t_{0})=5,000/N^{Y}$, $S_{0}^{Y}(t_{0})=1-5,000/N^{Y}$,
$S_{0}^{A}(t_{0})=1$, and $t_{0}=45\mathrm{d}$ (corresponding to
Mar 4, 2020 when Jan 1, 2020 corresponds to $t=0$). We find that
the temporal shape of the low-risk in-group reproduction number 
\begin{equation}
\mathcal{R}^{YY}(t)=\begin{cases}
2.25 & t<90\mathrm{d},\\
0.4 & 90\mathrm{d}\leq t<195\mathrm{d},\\
1.7 & 195\mathrm{d}\leq t<307\mathrm{d},\\
1.37 & 307\mathrm{d}\leq t<350\mathrm{d},\\
0.8 & 350\mathrm{d}\leq t
\end{cases}\label{eq:R_calib}
\end{equation}
replicates the daily fatalities in the age group 60+ during 2020 in
Germany (see Fig.~\ref{fig:Calibration-simulation}). We integrated
Eqs.~(\ref{eq:model_ODEs_begin}-\ref{eq:model_ODEs_end}) using Euler's
method for $t_{0}\leq t\leq365\mathrm{d}$ with incremental steps
of $\Delta t=0.25\mathrm{d}$. 

In order to explain differences in model incidence and observed incidence,
we have to assume that overall, 2 in 13 of the infecteds in group
$Y$ were officially reported, and that 10 in 18 infecteds in group
$A$ were reported. This corresponds to a dark factor of $1/\delta^{Y}=6.5$
in the low-risk group and $1/\delta^{A}=1.8$ in the high-risk group, resulting in a population-wide dark factor of $1/\delta=5.6$ which is 
comparable to the results of sero-prevalence studies in Germany \cite{streeck_infection_2020,santos-hovener_seroepidemiologische_2020}.

Note that official case data can be subject to fluctuations that are associated with reporting errors such as delay and erroneous entries, especially for case numbers that lie in the recent past. In order to test for the influence of such fluctuations, we varied the reproduction number $\mathcal R^{YY}(t)$ in the last quarter of 2020 by 5\% to find that such variations do not have any influence on our results, because they only influence the absolute cumulative value of symptomatic cases and deaths, but not their difference with respect to competing distribution protocols, which is our main observable.

\section{Stochastic simulations}

As defined above, the model is fully deterministic. However, large-scale
epidemic outbreaks are chaotic systems in the sense that environmental
influences as well as behavioral and societal feedback loops make
it impossible to predict the exact course of epidemics in the long-term
\cite{funk_spread_2009,funk_talk_2013,davies_effects_2020}. Rather,
several more general statements about the course of epidemics can
be made, for instance that decreasing case counts are usually met
with lifting restrictions and a return to normalcy, therefore causing
another resurgence. One may also expect that lockdowns are implemented
once case numbers rise again \cite{davies_effects_2020}. On shorter
time scales, environmental changes may impact the contact structure
\cite{schlosser_covid-19_2020} or viral transmission \cite{carlson_misconceptions_2020}.
In order to model the stochasticity of these influences, we transform
the deterministic model equations into stochastic differential equations
by defining a fluctuating low-risk reproduction number 
\begin{equation}
\mathrm{d}\mathcal{R}^{YY}=g(\mathcal{R}^{YY},t)\,\mathrm{d}t+h(\mathcal{R}^{YY},t)\,\mathrm{d}W_{t}\label{eq:sde_R}
\end{equation}
where $W_{t}$ is a Wiener process and the functions $g$ and $h$
are independent of the remaining system. As such, $d\mathcal{R}^{YY}$
is entirely decoupled from all other equations of motion. In order
to integrate the complete system numerically, it therefore suffices
to integrate Eq.~(\ref{eq:sde_R}) once using the Euler\textendash Maruyama
method with $\Delta t=0.25\mathrm{d}$. Afterwards, we can integrate
the model equations using Euler's method with the same $\Delta t$.
This is numerically equivalent to evaluating the whole system with
the Euler\textendash Maruyama method and enables us to evaluate the
outcome of different vaccination strategies with the same time series
$\mathcal{R}^{YY}(t)$. This means that even though we model a stochastic
system, the consequences of different vaccine rollout strategies can
be compared directly.

As initial condition, we choose the final value of Eq.~(\ref{eq:R_calib})
$\mathcal{R}^{YY}(t_{0})=0.8$ and redefine $t_{0}=0$ corresponding
to Jan 1, 2021. Furthermore we define
\begin{align*}
g(\mathcal{R}^{YY},t) & =-f\Big[\mathcal{R}^{YY}-\big(\hat{\mathcal{R}}+(t-t_{0})\varrho\big)\Big]\\
h(\mathcal{R}^{YY},t) & =\sqrt{2\zeta},
\end{align*}
i.e.~an Ornstein-Uhlenbeck process with drift. For all simulations,
we set $\zeta=s\sqrt{2f}$ and fix $s=0.002\mathrm{d}^{-1/2}$ and
$f=1/30$, which generates model incidence curves that roughly replicate
the autocorrelation time and standard deviation of the population-wide
7-day running average incidence curve in Germany in 2020. We also
employ reflective boundary conditions at $\mathcal{R}^{YY}=0$ and
$\mathcal{R}^{YY}=3$.

As explained above, we can expect an initially decreasing incidence
curve to resurge after a certain amount of time, causing a third wave
because restrictions are lifted while case counts are low and people
slowly return to their normal behavior. This leads to an increased
effective reproduction number. In order to obtain model curves that
replicate this behavior on a \emph{very} slow, a slow, and a fast
time-scale, we choose $\hat{\mathcal{R}}\in\left\{ 0.8,1.05,1.2\right\} $
and set a drift of $\varrho=0.005/\mathrm{d}$.

We dub reproduction number time series with $\hat{\mathcal{R}}=0.8$
as ``improving'' scenarios, those where $\hat{\mathcal{R}}=1.05$
as ``slow resurgence'' scenarios, and curves with $\hat{\mathcal{R}}=1.2$
as ``fast resurgence'' scenarios. We simulate $n_\mathrm{meas}=1000$ reproduction
number time series and use each to integrate the model equations for
all vaccination scenarios.

\section{Varying vaccine efficacy, transmission reduction, and adherence} 
\begin{figure*}
    \centering
    \includegraphics[width=0.5\textwidth]{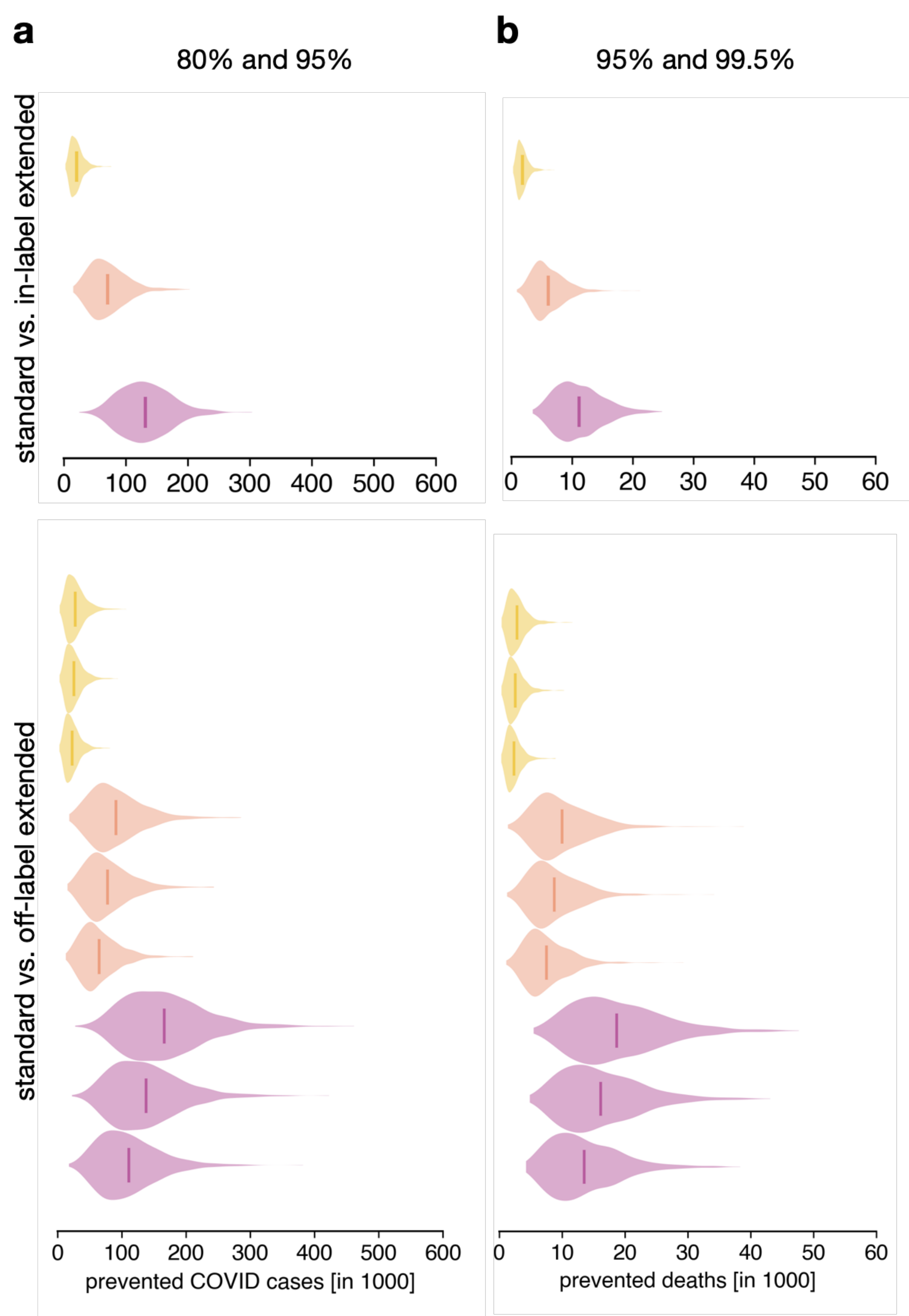}
    \caption{Varying vaccine efficacy against COVID-19 and COVID-19 related deaths for the vaccination protocols shown in Fig.~3a of the main text. (a) Prevented symptomatic COVID-19 cases with reduced efficacy against COVID-19 after the first dose, i.e. 80$\%$ after the first and 95 $\%$ efficacy after the second dose, respectively.
    (b) Prevented COVID-19-related deaths considering a reduced efficacy against death of 95$\%$ (first protection level) and 99.5 $\%$ (second protection level), respectively. Structure and color choice are analogous to Figs.~2-4 in the main text.
    Adjacent violins of the same color represent reductions in the adherence as described in the main text (see Fig.~4).}
    \label{fig:sens}
\end{figure*}
\begin{figure*}
    \centering
    \includegraphics[width=1\textwidth]{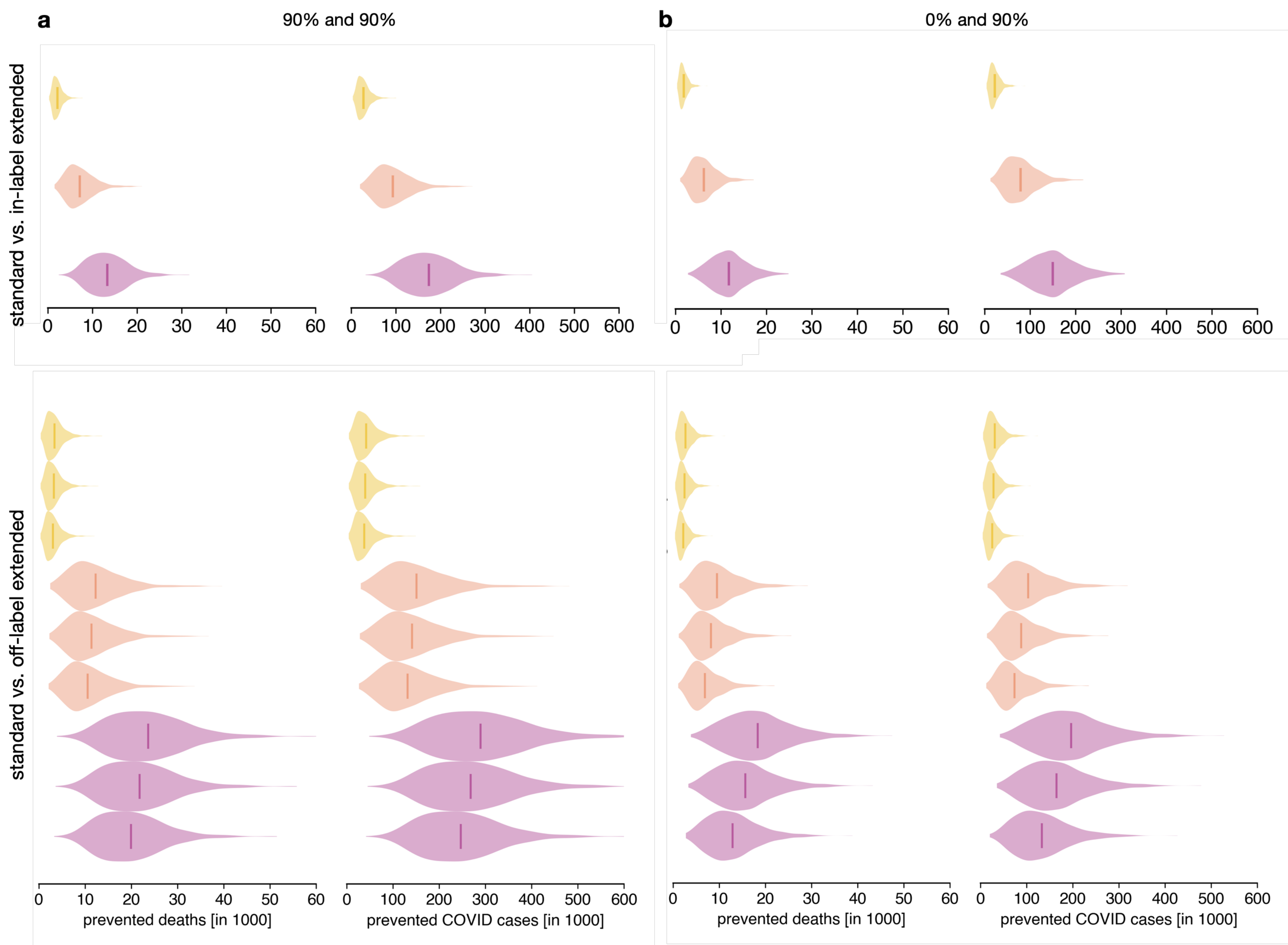}
    \caption{Prevented symptomatic COVID-19 cases and related deaths for varying transmissibility reduction for the vaccination protocols shown in Fig.~3a of the main text. (a)  $k_1=k_2=90\%$ (b) $k_1 = 0$ and $k_2 = 90\%$. Structure and color choice are analogous to Figs.~2-4 in the main text. Adjacent violins of the same color represent reductions in the adherence as described in the main text (see Fig.~4).}
    \label{fig:sens_trans}
\end{figure*}

\begin{figure*}
    \centering
    \includegraphics[width=0.5\textwidth]{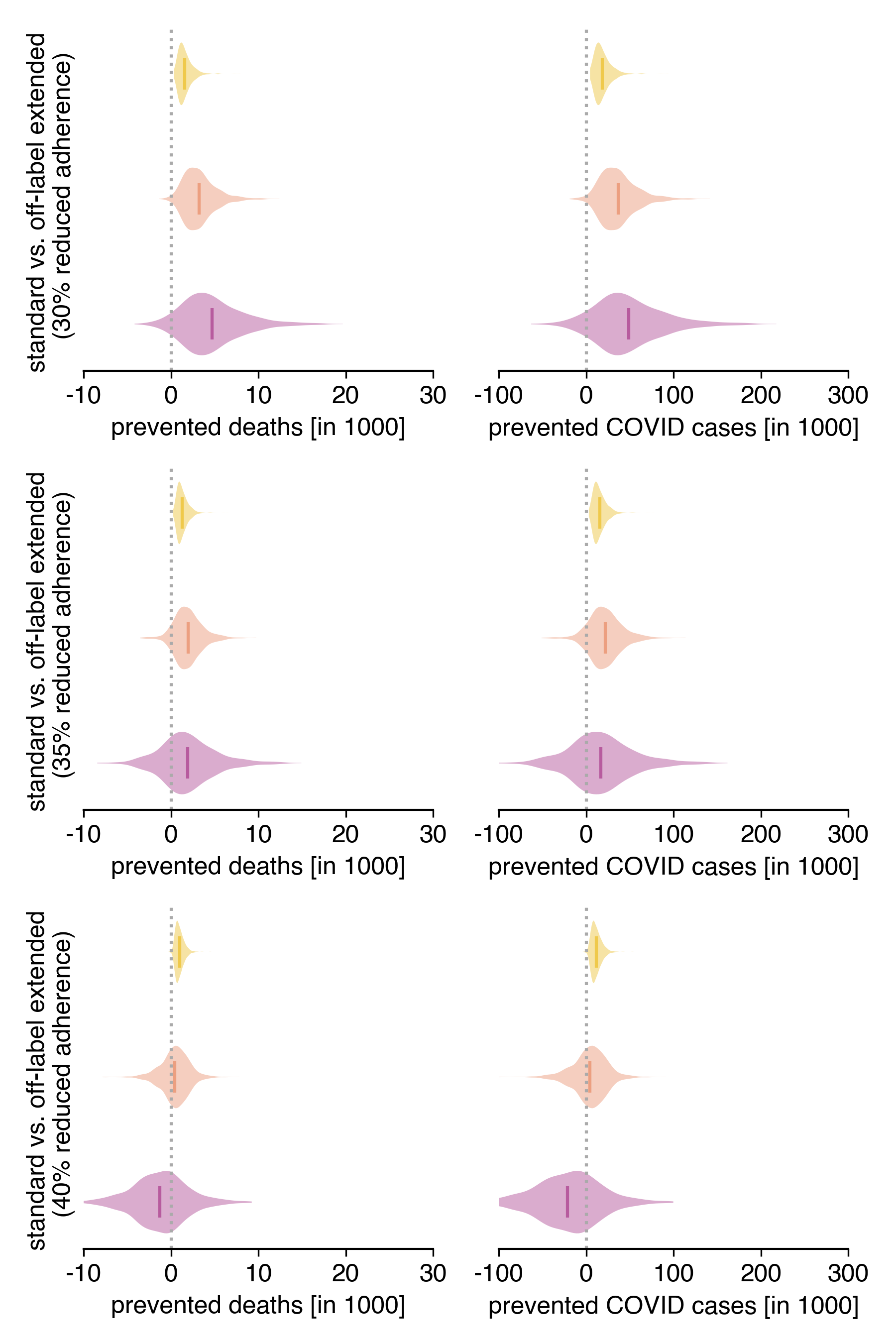}
    \caption{Prevented high-risk group deaths and symptomatic cases with decreasing adherence. \textbf{(Top row)} 30\% reduction (8.4 million high-risk individuals willing to get vaccinated, corresponds to a total of 55\% adherence). \textbf{(Middle row)} 35\% reduction (7.8 million high-risk individuals willing to get vaccinated, corresponds to a total of 51\% adherence). \textbf{(Bottom row)} 40\% reduction (7.2 million high-risk individuals willing to get vaccinated, corresponds to a total of 47\% adherence).}
    \label{fig:sens_adh}
\end{figure*}
In our baseline model configuration, we assume efficacy against COVID-19 to be $e_{\mathrm{COVID},1}=90\%$ after the first and  $e_{\mathrm{COVID},2}=95\%$ after the second dose as reported by the vaccine's respective phase 3 studies. Vaccine efficacy against death is assumed to be  $e_{\mathrm{COVID},1}=99.9\%$ and $e_{\mathrm{COVID},2}=99.99\%$ after reaching first and second protection levels, respectively. In order to test the robustness of our results regarding this baseline configuration, we performed additional sensitivity analyses where the efficacy of the first dose is substantially smaller than that provided by the second dose. In the following, we report the resulting relative changes of prevented deaths and COVID-19 cases with respect to the results obtained using the baseline configuration. Here, we only do analyses for vaccination distribution protocols shown in Fig.~3 of the main text.

First, we reduce first-dose efficacy against COVID-19 to $e_{\mathrm{COVID},1}=80\%$ while we keep the efficacy of the second dose fixed, therefore deliberately weakening the off-label protocol. A reduction from 90\% to 80\% efficacy implies an increase of 100\% in the probability to show symptoms after infection, hence representing a strong change. As expected, we find that such a strong reduction in first-dose efficacy leads to a reduction in the number of prevented ``high-risk'' COVID-19 cases $\Delta C^A$ while the positive net effect remains (i.e.~the number of prevented COVID-19 cases remains positive and large), see Fig.~\ref{fig:sens}a). We find a 17\% reduction in $\Delta C^A$ for the extended in-label protocol. Regarding the extended off-label protocol, $\Delta C^A$ is reduced by 26$\%$ for the ``improved'' and 35$\%$ for the ``fast resurgence'' scenario, respectively. Additionally reducing the adherence by 10\% yields a reduction of 28$\%$ (``improved'') and 42$\%$ (``fast resurgence'') in $\Delta C^A$, respectively. In summary, doubling the assumed probability of showing symptoms despite having received the first dose leads to a $\leq 50\%$ reduction in $\Delta C^A$, showing that even in this case, switching to the extended in-label or even to the extended off-label protocol would be beneficial.

Second, we decrease the efficacy against death to values of
$e_{\mathrm{death},1}=95\%$ for protection level 1 and
$e_{\mathrm{COVID},2}=99.5\%$ for protection level 2. This is a 50-fold
increase in the probability of death after infection as compared to the
baseline configuration. For the extended in-label protocol, such a
change leads to a $\approx 10\%$ reduction in prevented deaths.
Regarding the extended off-label protocol, this reduction efficacy would
yield an 11$\%$ to 15$\%$ reduction in prevented deaths (12$\%$ to
18$\%$ for a concurrent 10\% reduction in adherence). In summary, even for extremely low values of vaccine efficacy against death, the number of lives saved by changing protocols remains positive and in the four- to low five-digit range.

Reduction in transmissibility (i.e.~probability of transmission as quantified by viral shedding rates) for individuals that have been infected despite having received a vaccine is unknown at the time of writing. In the baseline configuration we have conservatively assumed that the vaccine causes no reduction in transmissibility for such individuals. Instead, in the following we want to analyze situations where the transmissibility reduction is large for both doses, as well as small for the first protection level and large for the second.

We find that implementing transmissibility reductions of $k_1=k_2=90\%$
lead to marginally improved results while reductions of $k_1 = 0$ and
$k_2 = 90\%$  led to a slight reduction of prevented COVID-19 cases and
deaths (see Fig.~\ref{fig:sens_trans}). For the in-label protocol,
values of $k_1=k_2=90\%$ would lead to an $8$\% increase of prevented
deaths $\Delta D^A$ and an 11\% increase in prevented COVID-19 cases
$\Delta C^A$ in the ``high-risk'' group (``improving'' scenario), and to
a 7\% ($\Delta D^A)$ and 9\% ($\Delta C^A$) increase for the ``fast
resurgence'' scenario, respectively. For the extended off-label protocol
these improvements range from 6\% to 13\% for both $\Delta D^A$ and
$\Delta C^A$ for all pandemic scenarios. We further find that the
negative consequences of a reduced adherence are not as pronounced in
this configuration as they are for the baseline configuration.

Regarding transmission reductions of $k_1 = 0$ (first protection level) and $k_2 = 90\%$ (second protection level) we find a decrease of prevented ``high-risk'' group deaths and cases by 7\% for all scenarios as compared to results using the baseline configuration (extended in-label protocol), see Fig~\ref{fig:sens_trans}. For the extended off-label protocol, the reduction in $\Delta C^A$ and $\Delta D^A$ ranges from 13\% to 25\% (max.~30\% for a concurrent 10\% reduction in adherence). Note that a situation where $k_1 = 0$ and $k_2 = 90\%$ represents an extreme and hence very unlikely scenario. Nevertheless, the number of saved lives and prevented cases in the ``high-risk'' group remains positive and in the four- to low five-digit range even in this extreme situation.

With decreased vaccine adherence in the high-risk population, the benefits of the extended off-label protocol are expected to vanish. For the baseline configuration, beginning at a 30\% relative adherence reduction, switching to the extended off-label protocol causes negative net effects, i.e.~negative numbers in the prevented deaths and cases (see Fig.~\ref{fig:sens_adh}). This corresponds to a willingness to get vaccinated of 55\% in the high-risk group (from 78.4\% pre-switch, i.e.~a 23.4\% reduction in the total willingness).

\section{Simulation code}
The model has been implemented based on the ``epipack'' modeling package \cite{maier_epipack_2021}. Implementation and analysis code is available online \cite{maier_devacc_2021}.
\section{Dose distribution scenarios}
We auto-generate distribution scenarios according to the following rules. We define the total number of high-risk individuals that will be vaccinated according to a chosen base adherence. We also define the number of high-priority low-risk individuals to be 7 million. Available doses are split between risk groups proportionally to the number of high-priority individuals in each group. Once all high-priority individuals have received two doses, all remaining doses are distributed to the low-risk group.

We further define one month to be 28 days long, in order to simplify distribution definitions in line with vaccine licensures. This implies that we integrate the model for 168 days, which according to our definitions amounts to 6 months.

Note that we round times to the nearest quarter of a day and in the tables, doses are rounded to the nearest integer.

All distribution scenarios that were generated according to these rules are displayed in Tabs.~\ref{tab:baseline_ficticious}-\ref{tab:off_label_extended_germany_reduced_adherence_strong}.

    \begin{table*}
    \centering
    \begin{tabular}{rrrrrrrrr}
\hline \hline
   $t_0$ [d] &   $t_0+T$ [d] &   $n^A_2$ &   $n^A_1$ &   $n^A_3$ &   $n^Y_2$ &   $n^Y_1$ &   $n^Y_3$ &   prot.~lev.~2nd dose \\
\hline
           0 &            84 &   8842105 &         0 &         0 &   5157894 &         0 &         0 &                     2 \\
          84 &           168 &   8842105 &         0 &         0 &   5157894 &         0 &         0 &                     2 \\
\hline
\end{tabular}
    \caption{Dose distribution according to the standard protocol for a supply situation of extreme scarcity (14 million doses in Q2), c.f.~Fig.~2 of the main text.\label{tab:baseline_ficticious}}
    \end{table*}

    \begin{table*}
    \centering
    \begin{tabular}{rrrrrrrrr}
\hline \hline
   $t_0$ [d] &   $t_0+T$ [d] &   $n^A_2$ &   $n^A_1$ &   $n^A_3$ &   $n^Y_2$ &   $n^Y_1$ &   $n^Y_3$ &   prot.~lev.~2nd dose \\
\hline
           0 &            42 &         0 &   4421052 &         0 &         0 &   2578947 &         0 &                     2 \\
          42 &            84 &         0 &         0 &   4421052 &         0 &         0 &   2578947 &                     2 \\
          84 &           126 &         0 &   4421052 &         0 &         0 &   2578947 &         0 &                     2 \\
         126 &           168 &         0 &         0 &   4421052 &         0 &         0 &   2578947 &                     2 \\
\hline
\end{tabular}
    \caption{Dose distribution according to the extended in-label protocol for a supply situation of extreme scarcity (14 million doses in Q2), c.f.~Fig.~2 of the main text.\label{tab:in_label_extended_ficticious}}
    \end{table*}

    \begin{table*}
    \centering
    \begin{tabular}{rrrrrrrrr}
\hline \hline
   $t_0$ [d] &   $t_0+T$ [d] &   $n^A_2$ &   $n^A_1$ &   $n^A_3$ &   $n^Y_2$ &   $n^Y_1$ &   $n^Y_3$ &   prot.~lev.~2nd dose \\
\hline
           0 &            84 &         0 &   8842105 &         0 &         0 &   5157894 &         0 &                     1 \\
          84 &           168 &         0 &         0 &   8842105 &         0 &         0 &   5157894 &                     1 \\
\hline
\end{tabular}
    \caption{Dose distribution according to the extended off-label protocol for a supply situation of extreme scarcity (14 million doses in Q2), c.f.~Fig.~2 of the main text.\label{tab:off_label_extended_ficticious}}
    \end{table*}

    \begin{table*}
    \centering
    \begin{tabular}{rrrrrrrrr}
\hline \hline
   $t_0$ [d] &   $t_0+T$ [d] &   $n^A_2$ &   $n^A_1$ &   $n^A_3$ &   $n^Y_2$ &   $n^Y_1$ &   $n^Y_3$ &   prot.~lev.~2nd dose \\
\hline
        0    &         84    &   8842105 &         0 &         0 &   5157894 &         0 &         0 &                     2 \\
       84    &        127.75 &  15131578 &         0 &         0 &   8826754 &         0 &         0 &                     2 \\
      127.75 &        168    &     26315 &         0 &         0 &  22015350 &         0 &         0 &                     2 \\
\hline
\end{tabular}
    \caption{Dose distribution according to the standard protocol for a supply situation of moderate scarcity (46 million doses in Q2), c.f.~Fig.~3 of the main text.\label{tab:baseline_germany}}
    \end{table*}

    \begin{table*}
    \centering
    \begin{tabular}{rrrrrrrrr}
\hline \hline
   $t_0$ [d] &   $t_0+T$ [d] &   $n^A_2$ &   $n^A_1$ &   $n^A_3$ &   $n^Y_2$ &   $n^Y_1$ &   $n^Y_3$ &   prot.~lev.~2nd dose \\
\hline
        0    &         42    &         0 &   4421052 &         0 &         0 &   2578947 &         0 &                     2 \\
       42    &         84    &         0 &         0 &   4421052 &         0 &         0 &   2578947 &                     2 \\
       84    &        105.75 &         0 &   7522556 &         0 &         0 &   4388157 &         0 &                     2 \\
      105.75 &        127.5  &         0 &         0 &   7522556 &         0 &         0 &   4388157 &                     2 \\
      127.5  &        147.75 &         0 &     28195 &         0 &         0 &  11061090 &         0 &                     2 \\
      147.75 &        168    &         0 &         0 &     28195 &         0 &         0 &  11061090 &                     2 \\
\hline
\end{tabular}
    \caption{Dose distribution according to the extended in-label protocol for a supply situation of moderate scarcity (46 million doses in Q2), c.f.~Fig.~3 of the main text.\label{tab:in_label_extended_germany}}
    \end{table*}

    \begin{table*}
    \centering
    \begin{tabular}{rrrrrrrrr}
\hline \hline
   $t_0$ [d] &   $t_0+T$ [d] &   $n^A_2$ &   $n^A_1$ &   $n^A_3$ &   $n^Y_2$ &   $n^Y_1$ &   $n^Y_3$ &   prot.~lev.~2nd dose \\
\hline
        0    &         84    &         0 &   8842105 &         0 &         0 &   5157894 &         0 &                     1 \\
       84    &         97    &         0 &   3127819 &   1368421 &         0 &   1824561 &    798245 &                     1 \\
       97    &        127.5  &         0 &         0 &  10548872 &         0 &         0 &   6153508 &                     1 \\
      127.5  &        147.75 &         0 &     30075 &         0 &         0 &  11059210 &         0 &                     2 \\
      147.75 &        168    &         0 &         0 &     30075 &         0 &         0 &  11059210 &                     2 \\
\hline
\end{tabular}
    \caption{Dose distribution according to the extended off-label protocol for a supply situation of moderate scarcity (46 million doses in Q2), c.f.~Fig.~3 of the main text.\label{tab:off_label_extended_germany}}
    \end{table*}

    \begin{table*}
    \centering
    \begin{tabular}{rrrrrrrrr}
\hline \hline
   $t_0$ [d] &   $t_0+T$ [d] &   $n^A_2$ &   $n^A_1$ &   $n^A_3$ &   $n^Y_2$ &   $n^Y_1$ &   $n^Y_3$ &   prot.~lev.~2nd dose \\
\hline
        0    &         84    &         0 &   8673913 &         0 &         0 &   5326086 &         0 &                     1 \\
       84    &         95.5  &         0 &   2714285 &   1187500 &         0 &   1666666 &    729166 &                     1 \\
       95.5  &        125.5  &         0 &         0 &  10178571 &         0 &         0 &   6250000 &                     1 \\
      125.5  &        146.75 &         0 &     11801 &         0 &         0 &  11625103 &         0 &                     2 \\
      146.75 &        168    &         0 &         0 &     11801 &         0 &         0 &  11625103 &                     2 \\
\hline
\end{tabular}
    \caption{Dose distribution according to the extended off-label protocol for a supply situation of moderate scarcity (46 million doses in Q2) and a relative adherence reduction of 5\%, c.f.~Fig.~4 of the main text. \label{tab:off_label_extended_germany_reduced_adherence_weak}}
    \end{table*}

    \begin{table*}
    \centering
    \begin{tabular}{rrrrrrrrr}
\hline \hline
   $t_0$ [d] &   $t_0+T$ [d] &   $n^A_2$ &   $n^A_1$ &   $n^A_3$ &   $n^Y_2$ &   $n^Y_1$ &   $n^Y_3$ &   prot.~lev.~2nd dose \\
\hline
        0    &         84    &         0 &   8494382 &         0 &         0 &   5505617 &         0 &                     1 \\
       84    &         93.5  &         0 &   2195826 &    960674 &         0 &   1423220 &    622659 &                     1 \\
       93.5  &        122.5  &         0 &         0 &   9635634 &         0 &         0 &   6245318 &                     1 \\
      122.5  &        145.25 &         0 &    109791 &         0 &         0 &  12348542 &         0 &                     2 \\
      145.25 &        168    &         0 &         0 &    109791 &         0 &         0 &  12348542 &                     2 \\
\hline
\end{tabular}
    \caption{Dose distribution according to the extended off-label protocol for a supply situation of moderate scarcity (46 million doses in Q2) and a relative adherence reduction of 10\%, c.f.~Fig.~4 of the main text.\label{tab:off_label_extended_germany_reduced_adherence_strong}}
    \end{table*}

\section{Full set of model equations}

In the following, we show all model equations for a single phase in
which the second, delayed dose only upholds the vaccine protection
of the first dose, rather than increasing it.

\begin{align}
\frac{d}{dt}S_{0}^{A} & =-\frac{S_{0}^{A}\left(2N^{A}\phi_{0}^{A}\left(S_{0}^{A}+\bar{S}_{0}^{A}\right)+\mu_{2}^{A}+2\nu_{1}^{A}\right)}{2N^{A}\left(S_{0}^{A}+\bar{S}_{0}^{A}\right)}\label{eq:model_ODEs_begin}\\
\frac{d}{dt}G_{0}^{A} & =-G_{0}^{A}\phi_{0}^{A}-\frac{G_{0}^{A}}{\tau_{M}}+\frac{S_{0}^{A}\nu_{1}^{A}}{N^{A}\left(S_{0}^{A}+\bar{S}_{0}^{A}\right)}\\
\frac{d}{dt}Z_{0}^{A} & =-Z_{0}^{A}\phi_{0}^{A}-\frac{Z_{0}^{A}}{\tau_{M}}+\frac{S_{0}^{A}\mu_{2}^{A}}{2N^{A}\left(S_{0}^{A}+\bar{S}_{0}^{A}\right)}\\
\frac{d}{dt}G_{1}^{A} & =-G_{1}^{A}\phi_{1}^{A}-\frac{G_{1}^{A}}{\tau_{M}}+\frac{W_{1}^{A}\nu_{2}^{A}}{N^{A}\left(W_{1}^{A}+\bar{W}_{1}^{A}\right)}
\end{align}

\begin{align}
\frac{d}{dt}W_{1}^{A} & =\frac{G_{0}^{A}}{\tau_{M}}-W_{1}^{A}\phi_{1}^{A}-\frac{W_{1}^{A}\nu_{2}^{A}}{N^{A}\left(W_{1}^{A}+\bar{W}_{1}^{A}\right)}\\
\frac{d}{dt}Z_{1}^{A} & =\frac{V_{1}^{A}-Z_{1}^{A}\phi_{1}^{A}\tau_{M}-Z_{1}^{A}}{\tau_{M}}\\
\frac{d}{dt}V_{1}^{A} & =\frac{-V_{1}^{A}\phi_{1}^{A}\tau_{M}-V_{1}^{A}+Z_{0}^{A}}{\tau_{M}}\\
\frac{d}{dt}K_{1}^{A} & =\frac{G_{1}^{A}}{\tau_{M}}-K_{1}^{A}\phi_{1}^{A}
\end{align}

\begin{align}
\frac{d}{dt}V_{2}^{A} & =-V_{2}^{A}\phi_{2}^{A}+\frac{Z_{1}^{A}}{\tau_{M}}\\
\frac{d}{dt}K_{2}^{A} & =-K_{2}^{A}\phi_{2}^{A}\\
\frac{d}{dt}\bar{S}_{0}^{A} & =\frac{N^{A}S_{0}^{A}\phi_{0}^{A}\left(S_{0}^{A}+\bar{S}_{0}^{A}\right)-\frac{\bar{S}_{0}^{A}\mu_{2}^{A}}{2}-\bar{S}_{0}^{A}\nu_{1}^{A}}{N^{A}\left(S_{0}^{A}+\bar{S}_{0}^{A}\right)}\\
\frac{d}{dt}\bar{G}_{0}^{A} & =G_{0}^{A}\phi_{0}^{A}-\frac{\bar{G}_{0}^{A}}{\tau_{M}}+\frac{\bar{S}_{0}^{A}\nu_{1}^{A}}{N^{A}\left(S_{0}^{A}+\bar{S}_{0}^{A}\right)}
\end{align}

\begin{align}
\frac{d}{dt}\bar{Z}_{0}^{A} & =Z_{0}^{A}\phi_{0}^{A}-\frac{\bar{Z}_{0}^{A}}{\tau_{M}}+\frac{\bar{S}_{0}^{A}\mu_{2}^{A}}{2N^{A}\left(S_{0}^{A}+\bar{S}_{0}^{A}\right)}\\
\frac{d}{dt}\bar{G}_{1}^{A} & =G_{1}^{A}\phi_{1}^{A}-\frac{\bar{G}_{1}^{A}}{\tau_{M}}+\frac{\bar{W}_{1}^{A}\nu_{2}^{A}}{N^{A}\left(W_{1}^{A}+\bar{W}_{1}^{A}\right)}\\
\frac{d}{dt}\bar{W}_{1}^{A} & =\frac{\bar{G}_{0}^{A}}{\tau_{M}}+W_{1}^{A}\phi_{1}^{A}-\frac{\bar{W}_{1}^{A}\nu_{2}^{A}}{N^{A}\left(W_{1}^{A}+\bar{W}_{1}^{A}\right)}\\
\frac{d}{dt}\bar{Z}_{1}^{A} & =\frac{\bar{V}_{1}^{A}+Z_{1}^{A}\phi_{1}^{A}\tau_{M}-\bar{Z}_{1}^{A}}{\tau_{M}}
\end{align}

\begin{align}
\frac{d}{dt}\bar{V}_{1}^{A} & =\frac{V_{1}^{A}\phi_{1}^{A}\tau_{M}-\bar{V}_{1}^{A}+\bar{Z}_{0}^{A}}{\tau_{M}}\\
\frac{d}{dt}\bar{K}_{1}^{A} & =\frac{\bar{G}_{1}^{A}}{\tau_{M}}+K_{1}^{A}\phi_{1}^{A}\\
\frac{d}{dt}\bar{V}_{2}^{A} & =V_{2}^{A}\phi_{2}^{A}+\frac{\bar{Z}_{1}^{A}}{\tau_{M}}\\
\frac{d}{dt}\bar{K}_{2}^{A} & =K_{2}^{A}\phi_{2}^{A}
\end{align}

\begin{align}
\frac{d}{dt}S_{0}^{Y} & =-\frac{S_{0}^{Y}\left(2N^{Y}\phi_{0}^{Y}\left(S_{0}^{Y}+\bar{S}_{0}^{Y}\right)+\mu_{2}^{Y}+2\nu_{1}^{Y}\right)}{2N^{Y}\left(S_{0}^{Y}+\bar{S}_{0}^{Y}\right)}\\
\frac{d}{dt}G_{0}^{Y} & =-G_{0}^{Y}\phi_{0}^{Y}-\frac{G_{0}^{Y}}{\tau_{M}}+\frac{S_{0}^{Y}\nu_{1}^{Y}}{N^{Y}\left(S_{0}^{Y}+\bar{S}_{0}^{Y}\right)}\\
\frac{d}{dt}Z_{0}^{Y} & =-Z_{0}^{Y}\phi_{0}^{Y}-\frac{Z_{0}^{Y}}{\tau_{M}}+\frac{S_{0}^{Y}\mu_{2}^{Y}}{2N^{Y}\left(S_{0}^{Y}+\bar{S}_{0}^{Y}\right)}\\
\frac{d}{dt}G_{1}^{Y} & =-G_{1}^{Y}\phi_{1}^{Y}-\frac{G_{1}^{Y}}{\tau_{M}}+\frac{W_{1}^{Y}\nu_{2}^{Y}}{N^{Y}\left(W_{1}^{Y}+\bar{W}_{1}^{Y}\right)}
\end{align}

\begin{align}
\frac{d}{dt}W_{1}^{Y} & =\frac{G_{0}^{Y}}{\tau_{M}}-W_{1}^{Y}\phi_{1}^{Y}-\frac{W_{1}^{Y}\nu_{2}^{Y}}{N^{Y}\left(W_{1}^{Y}+\bar{W}_{1}^{Y}\right)}\\
\frac{d}{dt}Z_{1}^{Y} & =\frac{V_{1}^{Y}-Z_{1}^{Y}\phi_{1}^{Y}\tau_{M}-Z_{1}^{Y}}{\tau_{M}}\\
\frac{d}{dt}V_{1}^{Y} & =\frac{-V_{1}^{Y}\phi_{1}^{Y}\tau_{M}-V_{1}^{Y}+Z_{0}^{Y}}{\tau_{M}}\\
\frac{d}{dt}K_{1}^{Y} & =\frac{G_{1}^{Y}}{\tau_{M}}-K_{1}^{Y}\phi_{1}^{Y}
\end{align}

\begin{align}
\frac{d}{dt}V_{2}^{Y} & =-V_{2}^{Y}\phi_{2}^{Y}+\frac{Z_{1}^{Y}}{\tau_{M}}\\
\frac{d}{dt}K_{2}^{Y} & =-K_{2}^{Y}\phi_{2}^{Y}\\
\frac{d}{dt}\bar{S}_{0}^{Y} & =\frac{N^{Y}S_{0}^{Y}\phi_{0}^{Y}\left(S_{0}^{Y}+\bar{S}_{0}^{Y}\right)-\frac{\bar{S}_{0}^{Y}\mu_{2}^{Y}}{2}-\bar{S}_{0}^{Y}\nu_{1}^{Y}}{N^{Y}\left(S_{0}^{Y}+\bar{S}_{0}^{Y}\right)}\\
\frac{d}{dt}\bar{G}_{0}^{Y} & =G_{0}^{Y}\phi_{0}^{Y}-\frac{\bar{G}_{0}^{Y}}{\tau_{M}}+\frac{\bar{S}_{0}^{Y}\nu_{1}^{Y}}{N^{Y}\left(S_{0}^{Y}+\bar{S}_{0}^{Y}\right)}
\end{align}

\begin{align}
\frac{d}{dt}\bar{Z}_{0}^{Y} & =Z_{0}^{Y}\phi_{0}^{Y}-\frac{\bar{Z}_{0}^{Y}}{\tau_{M}}+\frac{\bar{S}_{0}^{Y}\mu_{2}^{Y}}{2N^{Y}\left(S_{0}^{Y}+\bar{S}_{0}^{Y}\right)}\\
\frac{d}{dt}\bar{G}_{1}^{Y} & =G_{1}^{Y}\phi_{1}^{Y}-\frac{\bar{G}_{1}^{Y}}{\tau_{M}}+\frac{\bar{W}_{1}^{Y}\nu_{2}^{Y}}{N^{Y}\left(W_{1}^{Y}+\bar{W}_{1}^{Y}\right)}\\
\frac{d}{dt}\bar{W}_{1}^{Y} & =\frac{\bar{G}_{0}^{Y}}{\tau_{M}}+W_{1}^{Y}\phi_{1}^{Y}-\frac{\bar{W}_{1}^{Y}\nu_{2}^{Y}}{N^{Y}\left(W_{1}^{Y}+\bar{W}_{1}^{Y}\right)}\\
\frac{d}{dt}\bar{Z}_{1}^{Y} & =\frac{\bar{V}_{1}^{Y}+Z_{1}^{Y}\phi_{1}^{Y}\tau_{M}-\bar{Z}_{1}^{Y}}{\tau_{M}}
\end{align}

\begin{align}
\frac{d}{dt}\bar{V}_{1}^{Y} & =\frac{V_{1}^{Y}\phi_{1}^{Y}\tau_{M}-\bar{V}_{1}^{Y}+\bar{Z}_{0}^{Y}}{\tau_{M}}\\
\frac{d}{dt}\bar{K}_{1}^{Y} & =\frac{\bar{G}_{1}^{Y}}{\tau_{M}}+K_{1}^{Y}\phi_{1}^{Y}\\
\frac{d}{dt}\bar{V}_{2}^{Y} & =V_{2}^{Y}\phi_{2}^{Y}+\frac{\bar{Z}_{1}^{Y}}{\tau_{M}}\\
\frac{d}{dt}\bar{K}_{2}^{Y} & =K_{2}^{Y}\phi_{2}^{Y}
\end{align}

\begin{align}
\frac{d}{dt}E_{0}^{A} & =-E_{0}^{A}\alpha+G_{0}^{A}\phi_{0}^{A}+S_{0}^{A}\phi_{0}^{A}+Z_{0}^{A}\phi_{0}^{A}\\
\frac{d}{dt}I_{0}^{A} & =E_{0}^{A}\alpha-I_{0}^{A}\beta\\
\frac{d}{dt}R_{0}^{A} & =I_{0}^{A}\beta\\
\frac{d}{dt}D_{0}^{A} & =I_{0}^{A}\beta d_{0}^{A}
\end{align}

\begin{align}
\frac{d}{dt}C_{0}^{A} & =I_{0}^{A}\beta p_{0}^{A}\\
\frac{d}{dt}F_{0}^{A} & =E_{0}^{A}\alpha-\frac{F_{0}^{A}}{\tau_{R}}\\
\frac{d}{dt}X_{0}^{A} & =\frac{F_{0}^{A}}{\tau_{R}}\\
\frac{d}{dt}E_{1}^{A} & =-E_{1}^{A}\alpha+G_{1}^{A}\phi_{1}^{A}+K_{1}^{A}\phi_{1}^{A}+V_{1}^{A}\phi_{1}^{A}+W_{1}^{A}\phi_{1}^{A}+Z_{1}^{A}\phi_{1}^{A}
\end{align}

\begin{align}
\frac{d}{dt}I_{1}^{A} & =E_{1}^{A}\alpha-I_{1}^{A}\beta\\
\frac{d}{dt}R_{1}^{A} & =I_{1}^{A}\beta\\
\frac{d}{dt}D_{1}^{A} & =I_{1}^{A}\beta d_{1}^{A}\\
\frac{d}{dt}C_{1}^{A} & =I_{1}^{A}\beta p_{1}^{A}
\end{align}

\begin{align}
\frac{d}{dt}F_{1}^{A} & =E_{1}^{A}\alpha-\frac{F_{1}^{A}}{\tau_{R}}\\
\frac{d}{dt}X_{1}^{A} & =\frac{F_{1}^{A}}{\tau_{R}}\\
\frac{d}{dt}E_{2}^{A} & =-E_{2}^{A}\alpha+K_{2}^{A}\phi_{2}^{A}+V_{2}^{A}\phi_{2}^{A}\\
\frac{d}{dt}I_{2}^{A} & =E_{2}^{A}\alpha-I_{2}^{A}\beta
\end{align}

\begin{align}
\frac{d}{dt}R_{2}^{A} & =I_{2}^{A}\beta\\
\frac{d}{dt}D_{2}^{A} & =I_{2}^{A}\beta d_{2}^{A}\\
\frac{d}{dt}C_{2}^{A} & =I_{2}^{A}\beta p_{2}^{A}\\
\frac{d}{dt}F_{2}^{A} & =E_{2}^{A}\alpha-\frac{F_{2}^{A}}{\tau_{R}}
\end{align}

\begin{align}
\frac{d}{dt}X_{2}^{A} & =\frac{F_{2}^{A}}{\tau_{R}}\\
\frac{d}{dt}M_{1}^{A} & =-\nu_{1}^{A}\\
\frac{d}{dt}M_{2}^{A} & =-\mu_{2}^{A}\\
\frac{d}{dt}M_{3}^{A} & =-\nu_{2}^{A}
\end{align}

\begin{align}
\frac{d}{dt}E_{0}^{Y} & =-E_{0}^{Y}\alpha+G_{0}^{Y}\phi_{0}^{Y}+S_{0}^{Y}\phi_{0}^{Y}+Z_{0}^{Y}\phi_{0}^{Y}\\
\frac{d}{dt}I_{0}^{Y} & =E_{0}^{Y}\alpha-I_{0}^{Y}\beta\\
\frac{d}{dt}R_{0}^{Y} & =I_{0}^{Y}\beta\\
\frac{d}{dt}D_{0}^{Y} & =I_{0}^{Y}\beta d_{0}^{Y}
\end{align}

\begin{align}
\frac{d}{dt}C_{0}^{Y} & =I_{0}^{Y}\beta p_{0}^{Y}\\
\frac{d}{dt}F_{0}^{Y} & =E_{0}^{Y}\alpha-\frac{F_{0}^{Y}}{\tau_{R}}\\
\frac{d}{dt}X_{0}^{Y} & =\frac{F_{0}^{Y}}{\tau_{R}}\\
\frac{d}{dt}E_{1}^{Y} & =-E_{1}^{Y}\alpha+G_{1}^{Y}\phi_{1}^{Y}+K_{1}^{Y}\phi_{1}^{Y}+V_{1}^{Y}\phi_{1}^{Y}+W_{1}^{Y}\phi_{1}^{Y}+Z_{1}^{Y}\phi_{1}^{Y}
\end{align}

\begin{align}
\frac{d}{dt}I_{1}^{Y} & =E_{1}^{Y}\alpha-I_{1}^{Y}\beta\\
\frac{d}{dt}R_{1}^{Y} & =I_{1}^{Y}\beta\\
\frac{d}{dt}D_{1}^{Y} & =I_{1}^{Y}\beta d_{1}^{Y}\\
\frac{d}{dt}C_{1}^{Y} & =I_{1}^{Y}\beta p_{1}^{Y}
\end{align}

\begin{align}
\frac{d}{dt}F_{1}^{Y} & =E_{1}^{Y}\alpha-\frac{F_{1}^{Y}}{\tau_{R}}\\
\frac{d}{dt}X_{1}^{Y} & =\frac{F_{1}^{Y}}{\tau_{R}}\\
\frac{d}{dt}E_{2}^{Y} & =-E_{2}^{Y}\alpha+K_{2}^{Y}\phi_{2}^{Y}+V_{2}^{Y}\phi_{2}^{Y}\\
\frac{d}{dt}I_{2}^{Y} & =E_{2}^{Y}\alpha-I_{2}^{Y}\beta
\end{align}

\begin{align}
\frac{d}{dt}R_{2}^{Y} & =I_{2}^{Y}\beta\\
\frac{d}{dt}D_{2}^{Y} & =I_{2}^{Y}\beta d_{2}^{Y}\\
\frac{d}{dt}C_{2}^{Y} & =I_{2}^{Y}\beta p_{2}^{Y}\\
\frac{d}{dt}F_{2}^{Y} & =E_{2}^{Y}\alpha-\frac{F_{2}^{Y}}{\tau_{R}}
\end{align}

\begin{align}
\frac{d}{dt}X_{2}^{Y} & =\frac{F_{2}^{Y}}{\tau_{R}}\\
\frac{d}{dt}M_{1}^{Y} & =-\nu_{1}^{Y}\\
\frac{d}{dt}M_{2}^{Y} & =-\mu_{2}^{Y}\\
\frac{d}{dt}M_{3}^{Y} & =-\nu_{2}^{Y}\label{eq:model_ODEs_end}
\end{align}
\end{document}